\documentclass[twoside,twocolumn,9pt]{article}
\usepackage{extsizes}
\usepackage[super,sort&compress,comma]{natbib} 
\usepackage[version=3]{mhchem}
\usepackage[left=1.5cm, right=1.5cm, top=1.785cm, bottom=2.0cm]{geometry}
\usepackage{balance}
\usepackage{times}%,mathptmx}
\usepackage{sectsty}
\usepackage{lastpage}
\usepackage[format=plain,justification=justified,singlelinecheck=false,font={stretch=1.125,small,sf},labelfont=bf,labelsep=space]{caption}
\usepackage{float}
\usepackage{fancyhdr}
\usepackage{fnpos}
\usepackage[english]{babel}
\addto{\captionsenglish}{%
  
}
\usepackage{array}
\usepackage{droidsans}
\usepackage{charter}
\usepackage[T1]{fontenc}
\usepackage[usenames,dvipsnames]{xcolor}
\usepackage{setspace}
\usepackage[compact]{titlesec}
\usepackage{hyperref}
\usepackage[pdftex]{graphicx}
\DeclareGraphicsExtensions{.pdf,.png}
\graphicspath{{./figures/}}

\usepackage{grffile}

\usepackage{amsmath, amsthm, amssymb, amsfonts, amsbsy}
\usepackage{mathtools} % for dcases
\usepackage{bm}
\usepackage{mathrsfs}  

\renewcommand*{\v}[1]{\boldsymbol{#1}}
\newcommand\bnabla{\boldsymbol{\nabla}}
\newcommand\bcdot{\boldsymbol{\cdot}}
\renewcommand{\div}{\bnabla\bcdot} % divergence
\newcommand{\m}[1]{\v{#1}}

\renewcommand{\d}{\text{d}} % upright d
\newcommand{\id}{\, \d} % add spacing for integrals

\newcommand{\abs}[1]{\lvert #1 \rvert}

\DeclareMathOperator{\arctanh}{arctanh}
\DeclareMathOperator{\dilog}{dilog}

\usepackage[capitalise]{cleveref}
\crefformat{equation}{Eq. (#2#1#3)} % for simple equation
\Crefformat{equation}{Equation (#2#1#3)} % beginning of a sentence
\crefrangeformat{equation}{Eqs. (#3#1#4--#5\crefstripprefix{#1}{#2}#6)} % for range
\Crefrangeformat{equation}{Equations (#3#1#4--#5\crefstripprefix{#1}{#2}#6)} % beginning of a sentence
\crefmultiformat{equation}{Eqs. (#2#1#3)}{ and~(#2#1#3)}{, (#2#1#3)}{ and~(#2#1#3)} % for list with ``and''
\Crefmultiformat{equation}{Equations (#2#1#3)}{ and~(#2#1#3)}{, (#2#1#3)}{ and~(#2#1#3)} % beginning of a sentence

\usepackage{xcolor}
\definecolor{cream}{RGB}{222,217,201}

\begin{document}

\pagestyle{fancy}
\thispagestyle{plain}
\fancypagestyle{plain}{

%%%HEADER%%%
\fancyhead[C]{\includegraphics[width=18.5cm]{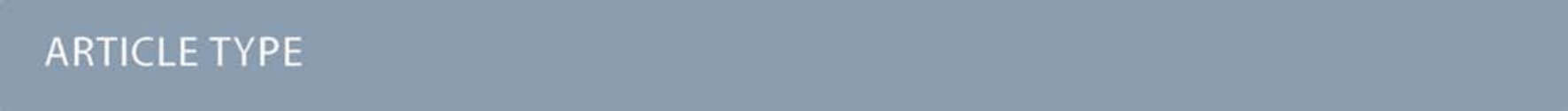}}
\fancyhead[L]{\hspace{0cm}\vspace{1.5cm}\includegraphics[height=30pt]{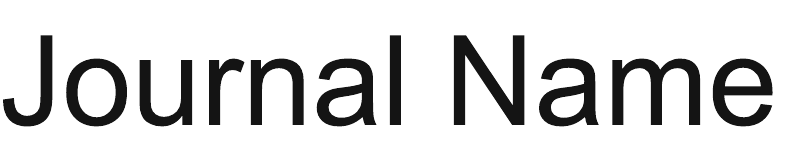}}
\fancyhead[R]{\hspace{0cm}\vspace{1.7cm}\includegraphics[height=55pt]{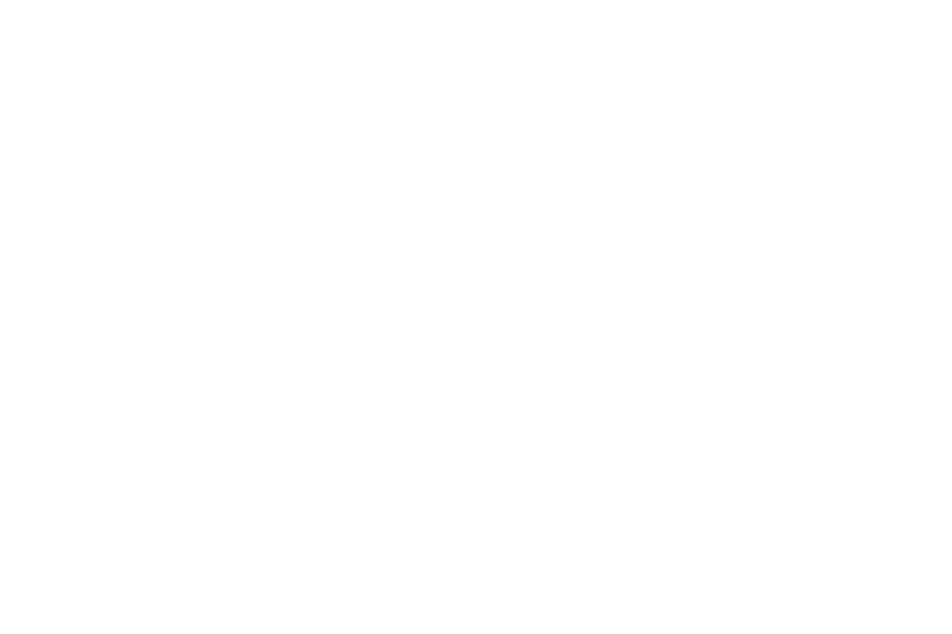}}
\renewcommand{\headrulewidth}{0pt}
}
%%%END OF HEADER%%%

%%%PAGE SETUP - Please do not change any commands within this section%%%
\makeFNbottom
\makeatletter
\renewcommand\LARGE{\@setfontsize\LARGE{15pt}{17}}
\renewcommand\Large{\@setfontsize\Large{12pt}{14}}
\renewcommand\large{\@setfontsize\large{10pt}{12}}
\renewcommand\footnotesize{\@setfontsize\footnotesize{7pt}{10}}
\makeatother

\renewcommand{\thefootnote}{\fnsymbol{footnote}}
\renewcommand\footnoterule{\vspace*{1pt}% 
\color{cream}\hrule width 3.5in height 0.4pt \color{black}\vspace*{5pt}} 
\setcounter{secnumdepth}{5}

\makeatletter 
\renewcommand\@biblabel[1]{#1}            
\renewcommand\@makefntext[1]% 
{\noindent\makebox[0pt][r]{\@thefnmark\,}#1}
\makeatother 
\renewcommand{\figurename}{\small{Fig.}~}
\sectionfont{\sffamily\Large}
\subsectionfont{\normalsize}
\subsubsectionfont{\bf}
\setstretch{1.125} %In particular, please do not alter this line.
\setlength{\skip\footins}{0.8cm}
\setlength{\footnotesep}{0.25cm}
\setlength{\jot}{10pt}
\titlespacing*{\section}{0pt}{4pt}{4pt}
\titlespacing*{\subsection}{0pt}{15pt}{1pt}
%%%END OF PAGE SETUP%%%

%%%FOOTER%%%
\fancyfoot{}
\fancyfoot[LO,RE]{\vspace{-7.1pt}\includegraphics[height=9pt]{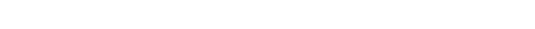}}
\fancyfoot[CO]{\vspace{-7.1pt}\hspace{13.2cm}\includegraphics{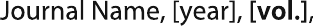}}
\fancyfoot[CE]{\vspace{-7.2pt}\hspace{-14.2cm}\includegraphics{head_foot/RF}}
\fancyfoot[RO]{\footnotesize{\sffamily{1--\pageref{LastPage} ~\textbar  \hspace{2pt}\thepage}}}
\fancyfoot[LE]{\footnotesize{\sffamily{\thepage~\textbar\hspace{3.45cm} 1--\pageref{LastPage}}}}
\fancyhead{}
\renewcommand{\headrulewidth}{0pt} 
\renewcommand{\footrulewidth}{0pt}
\setlength{\arrayrulewidth}{1pt}
\setlength{\columnsep}{6.5mm}
\setlength\bibsep{1pt}
%%%END OF FOOTER%%%

%%%FIGURE SETUP - please do not change any commands within this section%%%
\makeatletter 
\newlength{\figrulesep} 
\setlength{\figrulesep}{0.5\textfloatsep} 

\newcommand{\topfigrule}{\vspace*{-1pt}% 
\noindent{\color{cream}\rule[-\figrulesep]{\columnwidth}{1.5pt}} }

\newcommand{\botfigrule}{\vspace*{-2pt}% 
\noindent{\color{cream}\rule[\figrulesep]{\columnwidth}{1.5pt}} }

\newcommand{\dblfigrule}{\vspace*{-1pt}% 
\noindent{\color{cream}\rule[-\figrulesep]{\textwidth}{1.5pt}} }

\makeatother

%%%END OF FIGURE SETUP%%%

%%%TITLE, AUTHORS AND ABSTRACT%%%
\twocolumn[
  \begin{@twocolumnfalse}
\vspace{3cm}
\sffamily
\begin{tabular}{m{4.5cm} p{13.5cm} }

\includegraphics{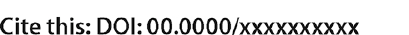} & \noindent\LARGE{\textbf{How many ways a cell can move: the modes of self-propulsion of an active drop}} \\%Article title goes here instead of the text "This is the title"
\vspace{0.3cm} & \vspace{0.3cm} \\

 & \noindent\large{Aurore Loisy \textit{$^{a}$}$^{\ast}$, Jens Eggers \textit{$^{a}$} and Tanniemola B. Liverpool \textit{$^{a}$}$^{\ddag}$} \\%Author names go here instead of "Full name", etc.

\includegraphics{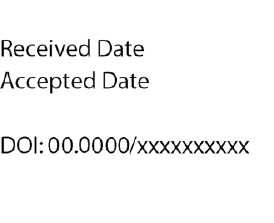} & \noindent\normalsize{Numerous physical models have been proposed to explain how cell motility emerges from internal activity, mostly focused on how crawling motion arises from internal processes. Here we offer a classification of self-propulsion mechanisms based on general physical principles, showing that crawling is not the only way for cells to move on a substrate. We consider a thin drop of active matter on a planar substrate and fully characterize its autonomous motion for all three possible sources of driving: (i) the stresses induced in the bulk by active components, which allow in particular tractionless motion, (ii) the self-propulsion of active components at the substrate, which gives rise to crawling motion, and (iii) a net capillary force, possibly self-generated, and coupled to internal activity. We determine travelling-wave solutions to the lubrication equations as a function of a dimensionless activity parameter for each mode of motion. Numerical simulations are used to characterize the drop motion over a wide range of activity magnitudes, and explicit analytical solutions in excellent agreement with the simulations are derived in the weak-activity regime.} \\

\end{tabular}

 \end{@twocolumnfalse} \vspace{0.6cm}

  ]
%%%END OF TITLE, AUTHORS AND ABSTRACT%%%

%%%FONT SETUP - please do not change any commands within this section
\renewcommand*\rmdefault{bch}\normalfont\upshape
\rmfamily
\section*{}
\vspace{-1cm}

%%%FOOTNOTES%%%

\footnotetext{\textit{$^{a}$~School of Mathematics, University of Bristol - Bristol BS8 1UG, UK.}}
\footnotetext{$^{\ast}$~E-mail: aurore.loisy@bristol.ac.uk.}
\footnotetext{$^{\ddag}$~E-mail: t.liverpool@bristol.ac.uk.}

% \footnotetext{\ddag~Additional footnotes to the title and authors can be included \textit{e.g.}\ `Present address:' or `These authors contributed equally to this work' as above using the symbols: \ddag, \textsection, and \P. Please place the appropriate symbol next to the author's name and include a \texttt{\textbackslash footnotetext} entry in the the correct place in the list.}

%%%END OF FOOTNOTES%%%

%%%MAIN TEXT%%%%

\section{Introduction}

To perform essential biological functions such as wound healing and immune response, but also in pathological processes such as cancer metastasis, eukaryotic cells adapt their mode of migration to the geometrical and physicochemical properties of their environment while relying on the same machinery, the actomyosin cytoskeleton \cite{Lammermann2009,Liu2015,Paluch2016}. 
In view of the complexity of cell motility, one may want to ask first: what are the physical requirements for autonomous motion, and what are the possible ways to move?
Here we answer these questions by taking a deformable drop of active matter (such as the cytoskeleton) and classifying the possible mechanisms for self-propulsion on a substrate. 

Motion on a hard surface is a particularly important class of motility, because it is the first step towards understanding the self-propulsion of cells in the tissue of multicellular organisms, and \textit{in vitro} experimental investigations of cell motility often involve the study of cells in contact with a solid substrate~\cite{Anderson1996,Verkhovsky1999,Yam2007,Keren2008,Lammermann2009}. % refs=exp papers
However, how such self-propulsion emerges from the components of living cells remains a subject of debate~\cite{Kruse2006,Keren2008,Shao2012,Blanch-Mercader2013,Callan-Jones2013,Recho2013,Tjhung2015,Khoromskaia2015}. % refs=cell motility on a surface or 1D model with friction

A minimal system to study motility is provided by a deformable drop of material with anisotropic components that consume energy (active matter) on a flat rigid surface~\cite{Sanchez2012,Tjhung2015,Khoromskaia2015}. % ref=active drop on a surface, exp and theory
For a drop of soft material to self-propel, two things are required: an asymmetry to give a direction of motion and a mechanical energy flux to provide the source of motion.  The asymmetry may be in the drop shape, resulting from an imbalance in surface tension, typically due to imposed chemical or thermal gradients which provide a non-zero flux leading to motion even for a passive drop~\cite{DeGennes2004book,Bonn2009,Brochard1989}. 
A drop of active matter, in contrast, generates fluxes and asymmetry all by itself due to energy input from its components~\cite{Ramaswamy2010,Marchetti2013,Saintillan2013,Prost2015,Julicher2018} that can cause the drop to move spontaneously~\cite{Sanchez2012,Hawkins2011,Tjhung2012,Tjhung2015,Giomi2014,Blanch-Mercader2013,Callan-Jones2013,Recho2013,Whitfield2016a}. % ref=spontaneous motion in various environments, exp and theory
Several studies have shown propulsion of active drops on a surface with a number of related models~\cite{Kruse2006,Shao2012,Ziebert2012,Tjhung2015,Khoromskaia2015,Trinschek2019arXiv}. % ref=motion on a surface, spontaneous or not
However the complexity of the underlying dynamics means identifying similarities and differences between them is difficult, leading to an ongoing debate about mechanisms.

The hydrodynamic theory of active matter provides a now well-accepted description of active liquids in terms of a limited number of coupled nonlinear governing equations for conserved fields and broken-symmetry fields \cite{Ramaswamy2010,Marchetti2013,Prost2015,Julicher2018}.
One way to study the problem of a moving active drop is through direct numerical simulations of those equations in a domain with moving boundaries  \cite{Tjhung2012,Tjhung2015,Giomi2014,Whitfield2014,Whitfield2016a,Marth2015}. While those provide valuable information, they are computationally expensive and they fail at providing a simple picture of the mechanisms at play. Another approach, which we shall follow here, takes advantage of the geometry of the problem: assuming that the drop is characterized by a small height-to-width ratio, one can use the disparity of length scales to reduce the full set of governing equations and boundary conditions to a single evolution equation much easier to analyze and comprehend. This framework, known as the lubrication (or long-wave) theory \cite{Oron1997,Craster2009}, has been exploited extensively for the study of thin films and droplets of passive nematic liquid crystals \cite{BenAmar2001,Cummings2004,Cummings2011,Lin2013a,Lin2013b,Crespo2017} and has recently been extended to active liquids with (nematic or polar) orientational order \cite{Sankararaman2009,Joanny2012,Khoromskaia2015,Kitavtsev2018,Trinschek2019arXiv}.
Prior work has been concerned with thin film stability \cite{Sankararaman2009}, dewetting \cite{Trinschek2019arXiv}, and drop spreading \cite{Joanny2012}. But to the best of our knowledge the question of motility has only been tackled superficially~\cite{Khoromskaia2015,Trinschek2019arXiv}, mostly due to the difficulty in obtaining a closed form for the evolution equation.

\begin{figure}
	\centering
	\includegraphics[width=0.99\linewidth]{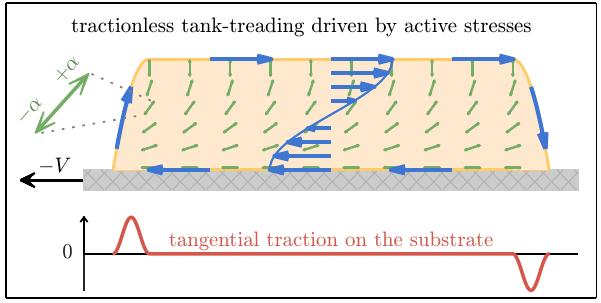} \\
	\includegraphics[width=0.99\linewidth]{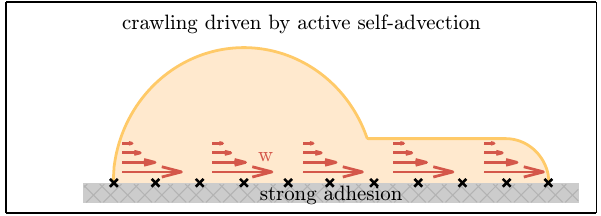} \\
	\includegraphics[width=0.99\linewidth]{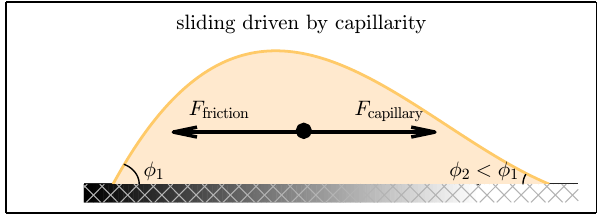}
	\caption{
	Classification of the modes of motion of an active drop.
	In tractionless tank-treading driven by active stresses, here drawn in the drop frame of reference, motion arises from the internal net flow (blue arrows) generated by active stresses ($\propto \alpha$), and is achieved without exerting any traction anywhere on the substrate except near the contact line.
	In crawling driven by self-advection, macroscopic motion arises from the self-advection ($\propto w$) of polarized active units, provided that adhesion with the substrate is strong enough to transmit momentum effectively. 
	In sliding driven by capillarity, the drop is pulled by a net capillary force (due to, e.g., an asymmetry in contact angles $\phi_1$ and $\phi_2$), the driving mechanism is an external or a self-generated gradient of surface tension or energy.
	\label{fig:scheme_intro} 
	}
\end{figure}

In this paper, we present a unifying description of a thin active drop on a planar substrate in terms of a single ODE. We show that its available modes of motion fall into three distinct classes which can be identified based on general principles, independent of the details of the model~(\cref{fig:scheme_intro}).

A first way to generate motion is through the extra stresses generated by the active components in the bulk. At the continuum scale, these active stresses yield an extra contribution to the stress tensor $\m{\sigma}^a = - \alpha \v{n} \v{n}$ where $\v{n}$ is the director (a unit vector that describes the local orientation of the active units). We show that the motion of a drop that originates from active stresses is controlled by the global topology of the director field, and can be achieved without exerting traction on the surface, a remarkable property which has been the subject of a recent communication \cite{Loisy2019b}.

The second possible source of motion is the self-advection term $w \v{n}$ that arises if the active units propel themselves at a speed $w$ along their own tangent. When coupled to strong enough friction with the substrate, self-advection allows a drop to ``crawl'' along the surface \cite{Kruse2006,Keren2008,Shao2010,Shao2012,Blanch-Mercader2013,Ziebert2013,Tjhung2015}. Crawling driven by self-advection encompasses much prior work on motile active drops on hard surfaces \cite{Kruse2006,Keren2008,Shao2010,Shao2012,Blanch-Mercader2013,Ziebert2013,Tjhung2015,Trinschek2019arXiv}, and is revisited here within our simple framework.

The third way to move is due to the action of a net capillary force, as would result from (possibly self-induced) thermal or chemical gradients. This mechanism has been exploited extensively to create self-propelled passive droplets \cite{DeGennes2004book,Bonn2009,Brochard1989,DeGennes1998,Chaudhury1992,Bain1994,DosSantos1995,Cira2015}, and here we address the effect of coupling it to internal activity.

\section{Model of a thin active drop \label{sec:model}}

\begin{figure}
	\centering
	\includegraphics[width=0.99\linewidth]{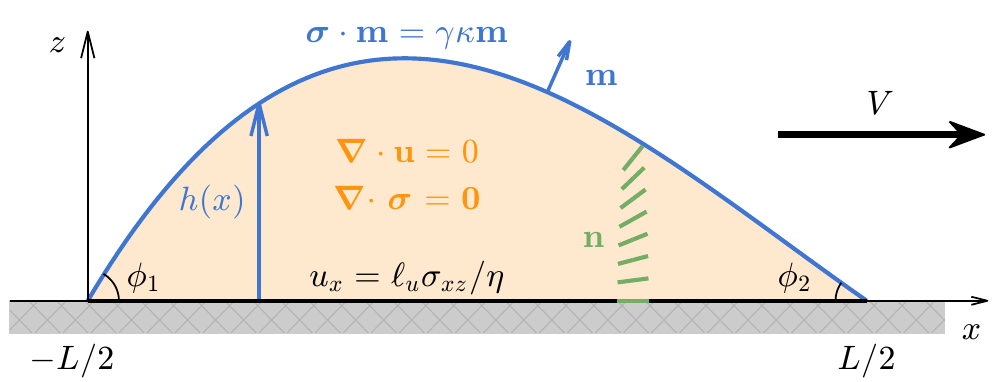}
	\caption{
	Model of a 2D drop of active fluid moving at velocity $V$ on a rigid surface. The fluid motion inside the drop is governed by the incompressible Stokes flow equations, with $\v{u}$ the velocity and $\m{\sigma}$ the stress tensor, which includes an active contribution $\m{\sigma}^a = - \alpha \v{n} \v{n}$ where $\v{n}$ is the director field. The mechanical interaction with the substrate is modeled by a partial slip boundary condition ($\ell_u$ is the slip length, $\eta$ is the viscosity) and a free surface boundary condition is applied at the interface ($\gamma$ is the surface tension coefficient and $\kappa$ is the curvature). The drop shape is described by the height function $h(x)$ on the domain $x \in [-L/2,L/2]$ where $L$ is the drop width. Contact angles $\phi_1$ and $\phi_2$ are prescribed on each side of the drop.
	\label{fig:scheme_drop_height_BC_annotated} 
	}
\end{figure}

Our model, illustrated in \cref{fig:scheme_drop_height_BC_annotated}, consists of a 2D drop of viscous, active, nematic liquid on a rigid substrate and confined by surface tension.
The director is strongly anchored at the boundaries, and the interaction of the liquid with the substrate is modelled by a partial slip boundary condition. The number density of active units is assumed uniform: motility induced by density gradients~\cite{Hawkins2011,Callan-Jones2013,Recho2013} is not considered here.
We further assume a drop geometry with a small height-to-width ratio and use the lubrication approximation to reduce the original problem to a nonlinear third-order ordinary differential equation for the drop shape which involves the drop velocity as an unknown constant and with prescribed contact angles as boundary conditions. It is obtained from the balance of activity, viscosity and surface tension in a regime where the director field minimizes the free energy (no backcoupling to the flow). In the following subsections we outline each of the ingredients that go into our model and analysis. The reader not interested in the details of the model and the derivation can find the thin drop problem we solve summarized in \cref{sec:reduced_problem}.

\subsection{Height equation \label{sec:height_equation}} 

We consider a drop moving on a substrate in the $x$-direction. At steady-state, the drop shape is described by the height function $h(x)$, and the constant drop velocity is denoted $V$ (both being unknown). In the co-moving frame of reference, the flux through a cross section must vanish. This reads
\begin{subequations}
\begin{equation}
\label{eq:unscaled_ODE}
  \int_0^h (u_x+w n_x - V) \; \d z = 0
\end{equation}
where $\v{u}$ is the fluid velocity inside the drop (with $\div \v{u} = 0$) and $w \v{n}$ describes the additional transport due to the self-advection at speed $w$ of active units whose orientations are characterised by a local orientation $\v{n}$ (a unit vector).

The height function, defined on the domain $x \in [-L/2, L/2$], must satisfy \cref{eq:unscaled_ODE} together with four boundary conditions at the contact lines:
\begin{equation}
\label{eq:unscaled_BC}
\begin{gathered}
 h(-\tfrac{L}{2}) = 0, \quad h(\tfrac{L}{2})=0, \\
 h'(-\tfrac{L}{2}) = \phi_1, \quad h'(\tfrac{L}{2}) = - \phi_2,
\end{gathered}
\end{equation}
where $\phi_{1,2}$ are the contact angles on each side of the drop.
The drop velocity $V$ enters as a constant which must be determined as part of the solution. The drop width $L$ is also unknown and is determined by the volume constraint
\begin{equation}
\label{eq:unscaled_volume_constraint}
 \int_{-L/2}^{L/2} h \; \d x = \Omega
\end{equation}
\end{subequations}
where $\Omega$ is the (prescribed) drop volume.
To close the problem described by \cref{eq:unscaled_ODE,eq:unscaled_BC,eq:unscaled_volume_constraint}, one must now determine an explicit expression of the integral on the left-hand-side of \cref{eq:unscaled_ODE} in terms of $h$.

\subsection{Self-advection of active units \label{sec:self_advection}}

The self-advection velocity $w \v{n}$ in \cref{eq:unscaled_ODE} accounts for the ability of polarized active components, such as motile bacteria or cytoskeletal filaments undergoing polymerization and treadmilling, to propel themselves along their own tangent. 
Such self-advection is confined close to the substrate, and to facilitate comparison with prior work we assume the same following functional form as in \cite{Tjhung2015}:
\begin{equation}
 w = w_0 \exp \left( - \frac{z}{\ell_w} \right)
 \label{eq:self_advection_expression}
\end{equation}
where $w_0$ is a characteristic self-advection speed and $\ell_w$ is the characteristic height over which the self-advection term decays in the direction normal to the substrate.

\subsection{Hydrodynamics of an active liquid \label{sec:hydrodynamics}}

The equations of motion for an active liquid are well-established \cite{Ramaswamy2010,Marchetti2013,Prost2015,Julicher2018}. 
Inside the drop, the velocity field is solution of the momentum conservation equation (neglecting inertia, see Appendix \ref{app:validity_regime}):
\begin{subequations}
\label{eq:momentum_conservation}
\begin{equation}
  \partial_j \sigma_{ij} = 0
  \label{eq:div_stress_zero}
\end{equation}
where $\sigma_{ij}$ is the stress tensor
\begin{equation}
	\label{eq:stress_definition}
	\sigma_{ij} = - p \delta_{ij} + \eta (\partial_j u_i + \partial_i u_j) + \sigma_{ij}^n + \sigma_{ij}^a
\end{equation}
and $p$ is the pressure, $\eta$ is the viscosity, $\sigma_{ij}^a$ is the contribution to the stress arising from activity, and $\sigma_{ij}^n$ is the contribution to the stress arising from its nematic elasticity \cite{deGennes1993book}.
The active stress reads \cite{Marchetti2013}
\begin{equation}
	\sigma_{ij}^a= -\alpha n_i n_j
\end{equation}
\end{subequations}
and is due to the forces exerted by the active units on the surrounding fluid. 
It can be derived from modeling active units as force dipoles \cite{Pedley1992} and subsequent coarse-graining. The magnitude of $\alpha$ is proportional to the strength of the force pair and the density of units, and the sign of $\alpha$ depends on whether the induced flow is extensile ($\alpha > 0$) or contractile ($\alpha < 0$).
For $\alpha=0$, one recovers the standard momentum balance for passive nematic liquid crystals. Since thin films and drops of passive nematics have been studied extensively (e.g. \cite{BenAmar2001,Cummings2004,Cummings2011,Lin2013a,Lin2013b,Crespo2017}), and since we are chiefly concerned here by $\alpha \neq 0$, we will first work in a regime where nematic stresses $\sigma_{ij}^n$ can be neglected (see Appendix \ref{app:validity_regime}). They will be included later on in Appendix \ref{app:effect_of_nematic_stresses}.

At the solid/liquid interface we use a partial slip boundary condition:
\begin{subequations}
\begin{equation}
	u_x = \frac{\ell_u \sigma_{xz}}{\eta}  \qquad \text{at $z=0$}
	\label{eq:BC_slip}
\end{equation}
where $\ell_u$ is a slip length (no-slip is obtained for $\ell_u = 0$). 
At the gas/liquid interface we use a free surface boundary condition:
\begin{equation}
	\m{\sigma} \bcdot \v{m} = \gamma \kappa \v{m} \qquad \text{at $z=h$}
	\label{eq:BC_surface_tension}
\end{equation}
\end{subequations}
where $\v{m}$ is the unit outward vector normal to the free surface, $\gamma$ is the uniform surface tension, and $\kappa = - \div \v{m}$ is the signed curvature.

The director $\v{n}=(\cos \theta, \sin \theta)$, which describes the coarse-grained orientation of the active units, is determined by minimizing the free energy of a nematic liquid crystal in the strong elastic limit~\cite{deGennes1993book}:
\begin{equation}
	\nabla^2 \theta = 0.
	\label{eq:director_eq}
\end{equation}
Hence the effect of the director on the flow is taken into account, but the back-coupling of the flow on the director is negligible in this regime (see Appendix \ref{app:validity_regime}).

As for boundary conditions, we assume strong anchoring (fixed angle relative to the surface orientation) at both the substrate and the free surface.
Restricting to situations where anchoring is either parallel or normal to the surfaces, and remarking that a rotation of $\v{n}$ by $\pi/2$ is equivalent to a change of sign of $\alpha$, we assume without loss of generality that the director is anchored parallel to the substrate:
\begin{subequations}
\label{eq:director_BC}
\begin{equation}
 \theta = 0 \qquad \text{at $z=0$}.
\end{equation}
At the free surface, we assume that the anchoring angle with respect to the surface tangent is $\omega \pi/2$ ($\omega \in \mathbb{Z}$), which reads
\begin{equation}
 \theta = \omega \frac{\pi}{2} + \arctan(h') \qquad \text{at $z=h$}.
\end{equation}
\end{subequations}

\subsection{Force and traction on the drop \label{sec:force_traction}}

Before going further it is useful to write down, without any simplifying assumptions, the force balance for a drop on a substrate. It reads (as shown in Appendix \ref{app:force_balance})
\begin{subequations}
\label{eq:force_balance}
\begin{equation}
 F_{\text{friction}} + F_{\text{capillary}} = 0
\end{equation}
with
\begin{align}
    F_{\text{capillary}} & = \gamma \left( \cos \phi_2 - \cos \phi_1 \right) \label{eq:Fcapilary}\\
    F_{\text{friction}} & = \int_{\text{substrate}} - \sigma_{xz} \rvert_{z=0} \id x \label{eq:Ffriction}
 \end{align}
\end{subequations}
where $F_{\text{capillary}}$ is a driving force due to an imbalance in surface energies (with $\phi_1$ and $\phi_2$ the contact angles on each side of the drop), and $F_{\text{friction}}$ is the opposing force (of frictional nature) exerted by substrate on the drop.
If the $\phi_1 \neq \phi_2$, $\abs{F_{\text{capillary}}}>0$: the drop is pulled by the net capillary force, and its velocity is determined by the balance with friction (leading to capillarity-driven sliding). If the contact angles are the same, $F_{\text{capillary}}=F_{\text{friction}}=0$: while a passive drop would necessarily remain static, this is not the case in the presence of activity (leading to tractionless tank-treading or crawling).

Besides, the mechanical interaction of the drop with the substrate can be characterized by the spatial distribution of the traction, the latter being defined as the \emph{local force per unit area} exerted by the substrate on the drop. 
The tangential component of the traction, denoted $\sigma_{\text{substrate/drop}}$, is 
\begin{equation}
 \sigma_{\text{substrate/drop}} = \v{e}_x \bcdot \m{\sigma} \bcdot (- \v{e}_z) \rvert_{z=0} = -\sigma_{xz} \rvert_{z=0}
\end{equation}
(in \cref{sec:results} we report instead $\sigma_{\text{drop/substrate}}=-\sigma_{\text{substrate/drop}}$ as this is what one would measure experimentally).

From \cref{eq:force_balance} one can remark that, for $F_{\text{capillary}}=0$, we necessarily have
\begin{equation}
	\int_{\text{substrate}} \sigma_{\text{substrate/drop}} \id x = 0
\end{equation}
but $\sigma_{\text{substrate/drop}}$ does not have to be identically zero. In other words, autonomous propulsion driven by active processes is necessarily force-free (in the sense that $F_{\text{capillary}}=F_{\text{friction}}=0$) but is not, in general, traction-free ($\sigma_{\text{drop/substrate}}$ is not zero everywhere).

\subsection{Lubrication approximation \label{sec:thin_film}}

We consider a geometry where the drop characteristic height $H$ is much smaller than its characteristic width $L$. We introduce a small parameter $\epsilon=H/L\ll 1$ and work in the framework of lubrication theory \cite{Oron1997,Craster2009}. 
Following the usual procedure (e.g., \cite{Cummings2011,Lin2013a,Crespo2017,Sankararaman2009,Kitavtsev2018,Joanny2012,Khoromskaia2015}), we rescale the coordinates and variables as follows: $\tilde{t}=(t U)/L$, $\tilde{x} = x / L$, $\tilde{z} = z/(\epsilon L)$, $\tilde{h} = h/(\epsilon L)$,  $\tilde{\ell}_{u,w} = \ell_{u,w}/(\epsilon L)$, $\tilde{u}_x = u_x / U$, $\tilde{u}_z = u_z / (\epsilon U)$, $\tilde{p} =  (p \epsilon^2 L) / (\eta U)$, $\tilde{\sigma}_{ij} = (\sigma_{ij} L)/(\eta U)$ where $U$ is a characteristic velocity scale in the $x$-direction for the internal flow.

We introduce several dimensionless groups that reflect the physics at play: $\mathcal{C} = \gamma/(\eta U)$ is an inverse capillary number which compares surface tension to viscous stresses, $\mathcal{A}=(\alpha L)/(\eta U)$ is the ratio of active stresses to viscous ones, and $\mathcal{W} = w_0/U$ controls the strength of self-advection compared to the internal fluid flow. 

At leading order in $\epsilon$, \cref{eq:director_eq} reduces to $\partial_{\tilde{z}}^2 \theta = 0$. Integrating twice and using the anchoring conditions [\cref{eq:director_BC}], we find the expression of the orientation field: 
\begin{equation}
 \theta = m \left( \frac{\omega \pi}{2} + \epsilon \tilde{h}' \right) \frac{\tilde{z}}{\tilde{h}}
\end{equation}
where $\omega$ is effectively a winding number which measures the number of quarter-turns of the director across the drop height, and where 
\begin{equation}
	m = \frac{h^2}{h^2 + \ell_\theta^2}
	\label{eq:m_regularization}
\end{equation}
is an \emph{ad hoc} regularizing function, borrowed from \cite{Cummings2011,Lin2013a}, and introduced to alleviate the conflict of strong anchoring conditions for $h \rightarrow 0$. 
Here $\ell_\theta$ is a characteristic small length scale such that for $h \gg \ell_\theta$, one retrieves the strong anchoring limit ($m=1$) and for $h \ll \ell_\theta$, the anchoring constraint is relaxed ($m=0$). 

Then, we have to distinguish between two situations: 
\begin{enumerate}
	\item $\omega \neq 0$ implies $\theta=O(1)$, therefore no rescaling is needed ($\tilde{\theta}=\theta$) and at leading order the director is not coupled to the drop shape;
	\item $\omega=0$, the director remains aligned with the bounding surfaces (deviations from the aligned state are due to deformations of the free interface), $\theta=O(\epsilon)$ so we rescale the director orientation as $\tilde{\theta}=\theta/\epsilon$.
\end{enumerate}
 
The expression of the (rescaled) director orientation is, at leading order in $\epsilon$,
\begin{equation}
  \tilde{\theta} = 
  \begin{dcases}
  	m \omega \pi \tilde{z} / (2 \tilde{h}) & \text{if $\omega \neq 0$,}\\
  	m \tilde{h}' \tilde{z} /\tilde{h} & \text{if $\omega = 0$.}
  \end{dcases}
\end{equation}
The derivation of the thin drop equation then proceeds as follows. The $z$-component of the Stokes flow equation [\cref{eq:div_stress_zero}] gives, at leading order in $\epsilon$, $\partial_{\tilde{z}} \tilde{p} = 0$, and using the normal component of the free surface boundary condition [\cref{eq:BC_surface_tension}] we find
\begin{equation}
	\tilde{p} = -\mathcal{C} \epsilon^3 \tilde{h}'' \qquad \qquad \text{for any $\omega$}.
\end{equation}
The $x$-component of \cref{eq:div_stress_zero} gives, at leading order, $\partial_{\tilde{z}} \tilde{\sigma}_{xz} = \partial_{\tilde{x}} \tilde{p}$.  This can be integrated once in $\tilde{z}$, and using the tangential component of \cref{eq:BC_surface_tension} we find the expression of the shear stress:
\begin{equation}
	\label{eq:lub_shear_stress}
  \tilde{\sigma}_{xz} =
  \begin{dcases}
  	- \mathcal{C} \epsilon^3 \tilde{h}''' (\tilde{z} - \tilde{h}) & \text{if $\omega \neq 0$,}\\
  	- \mathcal{C} \epsilon^3 \tilde{h}''' (\tilde{z} - \tilde{h}) - \mathcal{A} \epsilon^2 \tilde{h}' & \text{if $\omega = 0$.}
  \end{dcases}
\end{equation}
Substituting the definition of $\tilde{\sigma}_{xz}$ [\cref{eq:stress_definition}] into \cref{eq:lub_shear_stress} and integrating once in $\tilde{z}$ with the partial slip boundary condition [\cref{eq:BC_slip}] yields the parallel component of the fluid velocity:
\begin{subequations}
\label{eq;lub_parallel_velocity}
\begin{equation}
    \tilde{u}_x = \tilde{u}_x^c  + \tilde{u}_x^a
\end{equation}
with $\tilde{u}_x^c$ the capillary flow
\begin{equation}
  \tilde{u}_x^c = - \mathcal{C} \epsilon^3 \left( \frac{\tilde{z}^2}{2} - (\tilde{z}+\tilde{\ell}_u) \tilde{h} \right) \tilde{h}'''
\end{equation}
and $\tilde{u}_x^a$ the active flow
\begin{equation}
  \tilde{u}_x^a =
  \begin{dcases}
  	\mathcal{A} \epsilon \frac{(1 - \cos 2 \tilde{\theta})}{2 \pi \omega m} \tilde{h} & \text{if $\omega \neq 0$,}\\
  	\mathcal{A} \epsilon^2 \left( \frac{m \tilde{z}^2}{2} - (\tilde{z}+\tilde{\ell}_u) \tilde{h} \right) \frac{\tilde{h}'}{\tilde{h}} & \text{if $\omega = 0$.}
  \end{dcases}
\end{equation}
\end{subequations}
Averaging the flow over the drop height we find
\begin{subequations}
\label{eq:scaled_fluxes}
\begin{equation}
 \frac{1}{\tilde{h}} \int_0^{\tilde{h}} \tilde{u}_x^c \; \d \tilde{z} = \mathcal{C} \epsilon^3 \left( \frac{\tilde{h}}{3} + \tilde{\ell}_u \right) \tilde{h} \tilde{h}'''
\end{equation}
and
\begin{equation}
  \frac{1}{\tilde{h}} \int_0^{\tilde{h}} \tilde{u}_x^a \; \d \tilde{z} = 
  \begin{dcases}
  	\mathcal{A} \epsilon \frac{1}{2 \pi \omega m} \tilde{h} & \text{if $\omega \neq 0$,}\\
  	- \mathcal{A} \epsilon^2 \left( \frac{(3-m)\tilde{h}}{6} + \tilde{\ell}_u \right) \tilde{h}' & \text{if $\omega = 0$.}
  \end{dcases}
\end{equation}
Using the expression of $w$ given by \cref{eq:self_advection_expression} we can also write the mean flow due to self-advection
\begin{equation}
  \frac{1}{\tilde{h}} \int_0^{\tilde{h}} \tilde{w} \tilde{n}_x \; \d \tilde{z} = 
  \begin{dcases}
  	\text{not considered} & \text{if $\omega \neq 0$,} \\
  	\mathcal{W} \tilde{\ell}_w \frac{\left[ 1 - \exp \left( - \tilde{h}/\tilde{\ell}_w \right) \right]}{\tilde{h}} & \text{if $\omega = 0$.}
  \end{dcases}
\end{equation}
\end{subequations}
\Cref{eq:scaled_fluxes} closes \cref{eq:unscaled_ODE} which, in rescaled variables, can be written as
\begin{equation}
	\tilde{V} = \frac{1}{\tilde{h}} \int_0^{\tilde{h}} (\tilde{u}_x^c + \tilde{u}_x^a + \tilde{w} \tilde{n}_x ) \; \d \tilde{z}
\end{equation}

Note that we must have $\mathcal{C} \sim \epsilon^{-3}$ such that surface tension enters at leading order, $\mathcal{A} \sim \epsilon^{-1}$ for $\omega \neq 0$ and $\mathcal{A} \sim \epsilon^{-2}$ for $\omega=0$ such that active stresses play a role at leading order, and $\mathcal{W} \sim 1$ to have the effect of self-advection at leading order.

\subsection{Thin drop equation \label{sec:reduced_problem}}

To summarize, the steady-state shape $h$ of a thin active drop moving at constant (unknown and possibly zero) velocity $V$ along the substrate is the solution of a third-order nonlinear ODE. The form of this ODE depends on the winding number $\omega$, defined as the number of quarter-turns of the director across the drop height imposed by the anchoring boundary conditions. 

Introducing appropriate nondimensionalization and rescaling (denoted by a tilde), such that all rescaled quantities are $O(1)$, and defining
\begin{align}
	\tilde{\mathcal{V}} & = \frac{\eta V}{\gamma \epsilon^3}, \\
  \tilde{\mathcal{A}} & = 
 \begin{dcases}
    \frac{\alpha L}{2 \pi \omega \gamma \epsilon^2} & \text{if $\omega \neq 0$,}\\
    \frac{\alpha L}{\gamma \epsilon} & \text{if $\omega = 0$,}
\end{dcases} \\
 \tilde{\mathcal{W}} & = \frac{\eta w_0}{\gamma \epsilon^3}
\end{align}
we can write the problem as
\begin{subequations}
\label{eq:full_scaled_problem}
\begin{equation}
\label{eq:scaled_ODE}
\begin{split}
  & \left( \frac{\tilde{h}}{3} + \tilde{\ell}_u \right) \tilde{h} \tilde{h}''' + \tilde{\mathcal{A}} \tilde{f}^{\alpha}(\tilde{h}) + \tilde{\mathcal{W}} \tilde{f}^w(\tilde{h}) = \tilde{\mathcal{V}} \\
  \tilde{f}^{\alpha}(\tilde{h}) & = 
 \begin{dcases}
 \frac{\tilde{h}}{m} & \text{if $\omega \neq 0$,}\\
 - \left( \frac{(3-m) \tilde{h}}{6} + \tilde{\ell}_u \right) \tilde{h}'  & \text{if $\omega = 0$,}
\end{dcases} \\
  \tilde{f}^w(\tilde{h}) & = 
  \begin{dcases}
  	\text{not considered} & \text{if $\omega \neq 0$,}\\
  	\frac{\tilde{\ell}_w}{\tilde{h}} \left[ 1 - \exp \left( - \tilde{h}/\tilde{\ell}_w \right) \right] & \text{if $\omega = 0$,}
  \end{dcases}
\end{split}
\end{equation}
where $\tilde{\mathcal{V}}$ is the dimensionless rescaled drop velocity, to be determined as part of the solution, and $m$ is a regularizing function, defined by \cref{eq:m_regularization}, that relaxes the strong anchoring boundary conditions for $h \rightarrow 0$.
This ODE is supplemented by four boundary conditions
\begin{equation}
\label{eq:scaled_BC}
\begin{gathered}
 \tilde{h}(-\tfrac{\tilde{L}}{2}) = 0, \quad \tilde{h}(\tfrac{\tilde{L}}{2})=0, \\
 \tilde{h}'(-\tfrac{\tilde{L}}{2}) = \tilde{\phi}_1, \quad \tilde{h}'(\tfrac{\tilde{L}}{2}) = - \tilde{\phi}_2,
\end{gathered}
\end{equation}
where $\tilde{\phi}_{1,2}$ are the contact angles on each side of the drop, and where the drop width $\tilde{L}$ is determined from
\begin{equation}
 \int_{-\tilde{L}/2}^{\tilde{L}/2} \tilde{h} \; \d \tilde{x} = \tilde{\Omega}
 \label{eq:scaled_volume_constraint}
\end{equation}
\end{subequations}
where $\tilde{\Omega}$ is a prescribed drop volume.

To characterize the local mechanical interaction of the drop with the rigid surface, we also introduce 
\begin{equation}
  \tilde{\sigma}_{\mathrm{drop/substrate}} = \frac{L}{\gamma \epsilon^3} \sigma_{xz}\rvert_{z=0}
\end{equation}
which is the (rescaled dimensionless) local traction exerted by the drop on the surface in the x-direction. It can be expressed in terms of the local drop shape and reads:
\begin{equation}
\label{eq:definition_traction_surface}
  \tilde{\sigma}_{\mathrm{drop/substrate}} =
  \begin{dcases}
    \tilde{h} \tilde{h}''' & \text{if $\omega \neq 0$,}\\
    \tilde{h} \tilde{h}'''  - \tilde{\mathcal{A}} \tilde{h}' & \text{if $\omega = 0$.}
  \end{dcases}
\end{equation}

Finally the flow inside the drop is, at leading order, parallel to the wall. Redefining
\begin{equation}
	\tilde{u}_x = \frac{\eta u_x}{\gamma \epsilon^3}
\end{equation}
the rescaled fluid velocity is given as a function of the drop shape by
\begin{equation}
  \tilde{u}_x =
  \begin{dcases}
  	\tilde{u}_x^c + \tilde{\mathcal{A}} \frac{\tilde{h}}{m} \left[1 - \cos \left( \frac{\omega \pi \tilde{z}}{\tilde{h}}\right) \right] & \text{if $\omega \neq 0$,}\\
  	\tilde{u}_x^c + \tilde{\mathcal{A}} \left( \frac{m \tilde{z}^2}{2} - (\tilde{z}+\tilde{\ell}_u) \tilde{h} \right) \frac{\tilde{h}'}{\tilde{h}}  & \text{if $\omega = 0$,}
  \end{dcases}
\end{equation}
where $\tilde{u}_c$ is the usual capillary parabolic flow
\begin{equation}
 \tilde{u}_x^c = - \left( \frac{\tilde{z}^2}{2} - (\tilde{z}+\tilde{\ell}_u) \tilde{h} \right) \tilde{h}'''.
\end{equation}

\subsection{Numerical methods and parameters}

Stable solutions to \cref{eq:scaled_ODE,eq:scaled_BC,eq:scaled_volume_constraint} and presented in \cref{sec:results} were obtained numerically as steady solutions to the time-dependent problem (presented in \cref{app:height_equation} and given by \cref{eq:height_eq_PDE}) in the thin drop approximation.

Our time integration algorithm is based on a Crank-Nicolson scheme with adaptive time-stepping. For space discretization, we use second-order finite difference schemes on a uniform grid. At each time step, the resulting nonlinear system of equations was solved using the Matlab nonlinear system solver. The solution was advanced in time until the steady-state was reached, corresponding to the sought-after travelling-wave solution. 

Numerical parameters used in the simulations are summarized in \cref{tab:numerical_parameters}. The volume (surface area) of the drop was kept constant across all the simulations and set to $\tilde{\Omega}=1$. 

\begin{table}[!h]
\begin{tabular*}{0.48\textwidth}{@{\extracolsep{\fill}}lccc}
  \hline
			& \multicolumn{3}{c}{self-propulsion driven by} \\
                        & active stresses & self-advection & capillarity \\
  \hline
  $N_{\text{grid}}$     & 800 & 400 & 200 \\
  $\tilde{\Omega}$	& 1 & 1 & 1 \\
  $\tilde{\ell}_u$	& 0.05 & 0.01 & 0.05 \\	
  $\tilde{\ell}_\theta$	& 0.05 & 0.01 & 0.05 \\	
  $\tilde{\phi}_1$	& 1 & 1 & 10 \\
  $\tilde{\phi}_2$	& 1 & 1 & 5 \\
  \hline
\end{tabular*}
\caption{Numerical parameters used in the simulations (unless mentioned otherwise): number of grid points ($N_{\text{grid}}$), drop volume ($\tilde{\Omega}$), slip length ($\tilde{\ell}_u$), characteristic thickness for strong anchoring relaxation ($\tilde{\ell}_\theta$, set equal to $\tilde{\ell}_u$), and contact angles ($\tilde{\phi}_1$ and $\tilde{\phi}_2$).
\label{tab:numerical_parameters}}
\end{table}

\section{Results \label{sec:results}}

Three distinct driving mechanisms (active stresses, self-advection, and capillary forces due to different contact angles) are embedded in \cref{eq:scaled_ODE,eq:scaled_BC,eq:scaled_volume_constraint}, leading to the three modes of motion summarized in \cref{tab:modes_of_motion_parameters} and that we will analyze separately in the following.

\begin{table}[!h]
\begin{tabular*}{0.48\textwidth}{@{\extracolsep{\fill}}cccc}
	\hline
			& \multicolumn{3}{c}{self-propulsion driven by} \\
                        & active stresses & self-advection & capillarity \\
  \hline
  $\omega$		& $\neq 0$ & 0 & 0 \\
  $\tilde{\mathcal{A}}$ & $\neq 0$ & 0 & 0 and $\neq 0$ \\
  $\tilde{\mathcal{W}}$ & 0 & $\neq 0$ & 0 \\ 
  $\phi_2-\phi_1$	& 0 & 0 & $\neq 0$ \\
  \hline
\end{tabular*}
\caption{The three basic modes of motion: (i) tractionless tank-treading driven by active stresses ($\propto \tilde{\mathcal{A}}$) and controlled by the winding number $\omega$ (if $\omega=0$ the drop is static), (ii) crawling driven by self-advection ($\propto \tilde{\mathcal{W}}$), (iii) sliding driven by a capillary force $F_{\text{capillary}}=\gamma (\cos \phi_2 - \cos \phi_1)$ and possibly modulated by activity. 
\label{tab:modes_of_motion_parameters}}
\end{table}

\subsection{Self-propulsion driven by active stresses \label{sec:TTT}} 

\begin{figure*}
    \flushleft
    (a) \\[-1em]
    \centering
    \includegraphics[height=3.1cm]{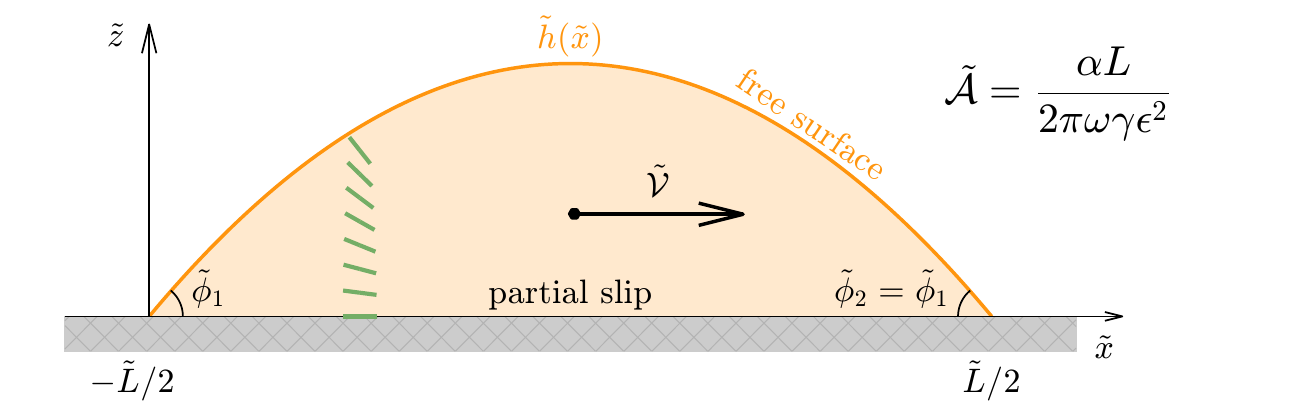}
    \includegraphics[height=3.1cm]{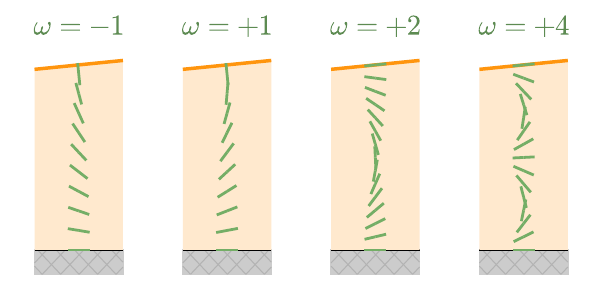} \\
    \vspace{-0.5em}
    \flushleft
    (b) \\[-0.3em]
    \centering
    \includegraphics[width=0.99\linewidth]{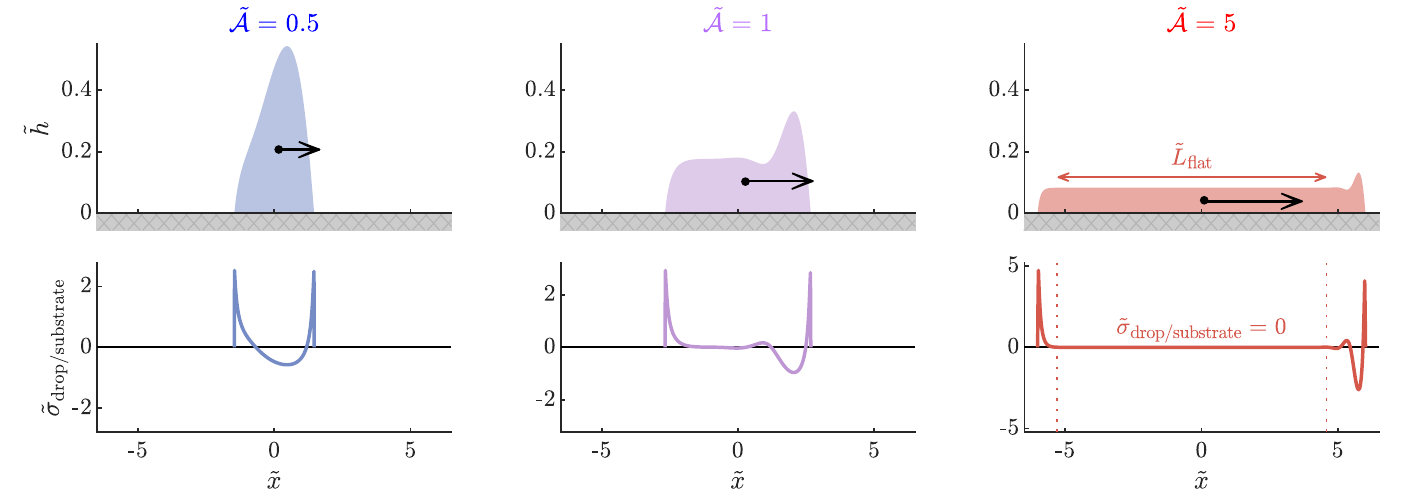} \\
    \vspace{-0.5em}
    \flushleft
    (c) \hspace{11.8cm} (d) \\[-0.3em]
    \centering
    \includegraphics[width=0.66\linewidth]{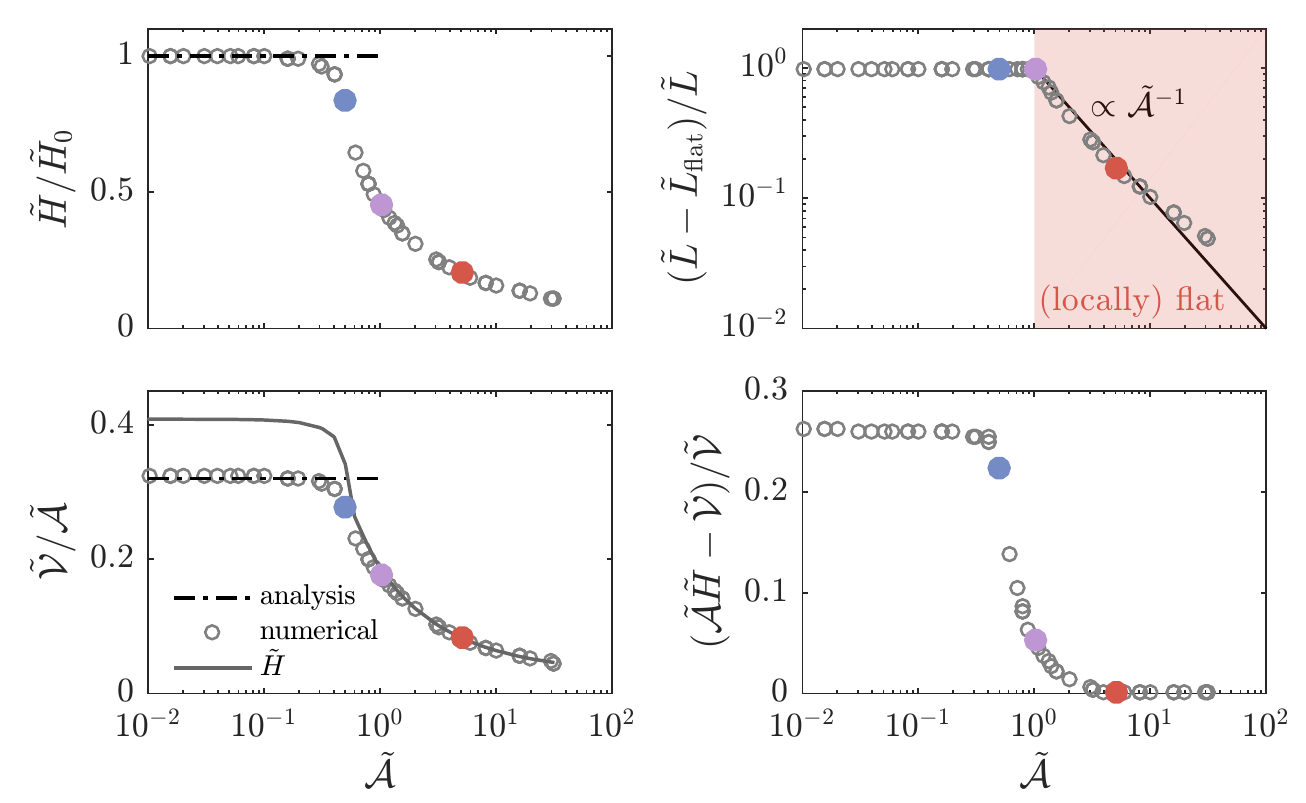} 
    \includegraphics[width=0.33\linewidth]{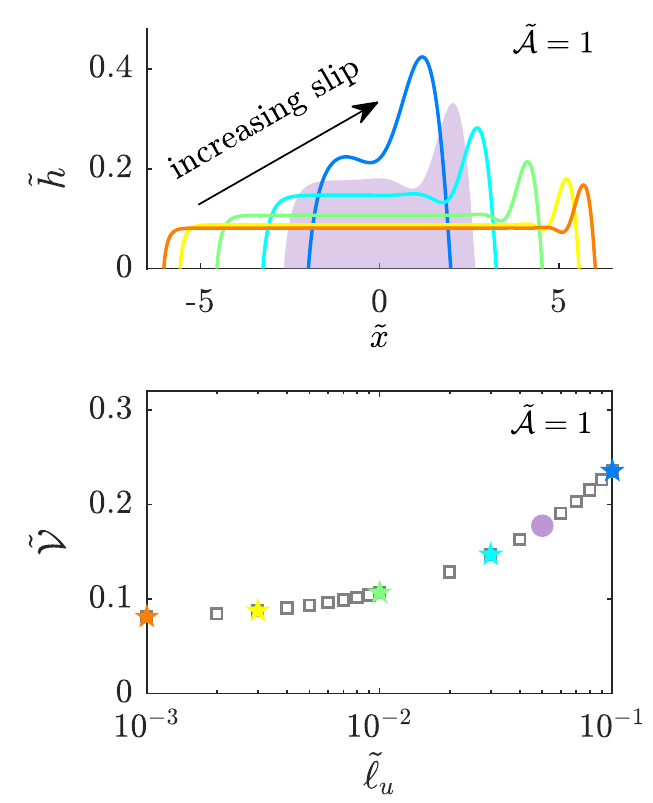} \\
    \flushleft
    (e) \\[-1em]
    \centering
    \includegraphics[width=0.97\linewidth]{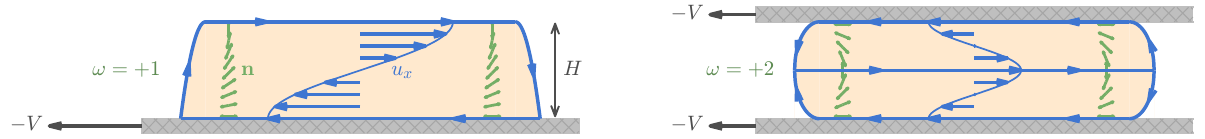}
    \caption{
	Self-propulsion driven by active stresses: the motion of a drop endowed with active stresses is controlled by the global topology of the director field and can be achieved without exerting traction locally on the surface.
	(a) Model of a thin active drop with active stresses and a winded director. The drop shape and velocity are controlled by $\tilde{\mathcal{A}} =(\alpha L) / ( 2 \pi \omega \gamma \epsilon^2)$ where $\omega$ is the winding number.
	(b) Numerical profiles of the drop shape and tangential traction exerted on the substrate.
	(c) Effect of $\tilde{\mathcal{A}}$ on the drop shape and velocity, $\tilde{H}$ is the mean height and the analytical solution for $\tilde{\mathcal{V}}$ is given by \cref{eq:TTT_pert_an}.
	(d) Effect of slip on the drop shape and velocity ($\mathcal{\tilde{A}}=1$).
	(e) Sketch of a tractionless flat drop moving at velocity $V=(\alpha H)/(2 \pi \omega \eta)$: director and velocity fields in the co-moving frame of reference (solution for $\omega=2$ is also valid for a drop confined between two walls).
	Colored symbols in (b-d) mark corresponding state points across panels.
	\label{fig:TractionlessTankTreading} 
	}
\end{figure*}

We consider the motion of a drop arising solely from active stresses ($\tilde{\mathcal{W}}=0$, $\tilde{\phi}_1 = \tilde{\phi}_2$). We found that if $\omega=0$, the drop is static (we comment on this at the end of this subsection). Therefore we assume $\omega \neq 0$, that is, we enforce a winding of the director through anchoring conditions at the bounding surfaces [\cref{fig:TractionlessTankTreading}(a)]. The governing ODE for the drop shape reduces to
\begin{equation}
\label{eq:TTT}
   \left( \frac{\tilde{h}}{3} + \tilde{\ell}_u \right) \tilde{h} \tilde{h}''' + \tilde{\mathcal{A}} \frac{\tilde{h}}{m} = \tilde{\mathcal{V}}.
\end{equation}
One can readily see that the drop shape and velocity are controlled by the dimensionless parameter $\tilde{\mathcal{A}} = (\alpha L)/(2 \pi \omega \gamma \epsilon^2)$, where $\alpha$ and $\omega$ can be of either sign. 

Since $\tilde{\mathcal{A}}(\alpha,\omega)=\tilde{\mathcal{A}}(-\alpha,-\omega)$, changing the direction in which the director winds (from counter-clockwise to clockwise) is equivalent to changing the sign of activity (from extensile to contractile). It is also interesting to note that if $\{\tilde{h}(\tilde{x}),\mathcal{V}\}$ is a solution for $\tilde{\mathcal{A}}$ then $\{\tilde{h}(-\tilde{x}),-\tilde{\mathcal{V}}\}$ is a solution for $-\tilde{\mathcal{A}}$: reversing the sign of $\tilde{\mathcal{A}}$ simply reverses the direction of motion. Therefore in the following we will only consider $\tilde{\mathcal{A}} \geqslant 0$.

The evolution of the drop shape and velocity with $\tilde{\mathcal{A}}$ is shown in \cref{fig:TractionlessTankTreading}(b,c). Overall, the drop becomes thinner and faster as activity increases. Solutions are however qualitatively different at low and high $\tilde{\mathcal{A}}$.

In the limit of small $\tilde{\mathcal{A}}$, the drop shape is close to a parabola (the equilibrium shape for a passive drop), and its velocity can be computed analytically at linear order in $\tilde{\mathcal{A}}$ (Appendix \ref{app:TTT_perturbation_solution}):
\begin{equation}
\label{eq:TTT_pert_an}
 \tilde{\mathcal{V}} = \tilde{\mathcal{A}} \frac{\tilde{\phi} \tilde{L}_0}{4} \frac{\sqrt{\tilde{b}} - (\tilde{b}-1) \arctanh(1/\sqrt{\tilde{b}})}{\arctanh(1/\sqrt{\tilde{b}})}
\end{equation}
where $\tilde{b}=1 + 12 \tilde{\ell}_u/(\tilde{L}_0 \tilde{\phi})$ and $\tilde{L}_0= \sqrt{6 \tilde{\Omega}/\tilde{\phi}}$. The first correction for the drop width is quadratic, so at linear order $\tilde{L}=\tilde{L}_0=\sqrt{6 \tilde{\Omega}/\tilde{\phi}}$ and the mean drop height is $\tilde{H}=\tilde{H}_0=\tilde{\Omega}/\tilde{L}_0$ where $\tilde{\Omega}$ is the drop volume (kept constant across simulations). Comparison to the numerical solution (dash-dotted lines in \cref{fig:TractionlessTankTreading}(c) (left panels) is excellent and shows that this solution remains valid up to $\tilde{\mathcal{A}} \approx 0.1$.

For $\tilde{\mathcal{A}} \geqslant 1$, the drop is locally flat (the numerical value of this threshold depends on $\tilde{\ell}_u$ and $\phi$, so the fact it is unity here is coincidental). The extent of the flat region rapidly increases with $\tilde{\mathcal{A}}$: analysis of numerical data indicates that, for $\tilde{\mathcal{A}} \geqslant 1$, the fraction of the drop which is \emph{not flat} first decreases as $\tilde{\mathcal{A}}^{-1}$. For $\tilde{\mathcal{A}} \gtrsim 10$, more than $90 \%$ of the drop is bounded by a flat free surface [\cref{fig:TractionlessTankTreading}(c), top right panel].
In this regime the drop velocity is exactly given by $\tilde{\mathcal{V}} = \tilde{\mathcal{A}} \tilde{h}_{\textrm{flat}}$ where $\tilde{h}_{\textrm{flat}}$ is the height of the flat region. 
In practice, the drop velocity is well approximated by [\cref{fig:TractionlessTankTreading}(c), bottom panels]
\begin{equation}
    \label{eq:flat_drop_velocity}
  \tilde{\mathcal{V}} = \tilde{\mathcal{A}} \tilde{H} %\qquad \text{for $\tilde{\mathcal{A}} \geqslant 1$}
\end{equation}
with $\tilde{H}$ the mean drop height: the error on $\tilde{\mathcal{V}}$ is less than 5 \% for $\tilde{\mathcal{A}} \geqslant 1$ and goes to zero as $\tilde{\mathcal{A}} \rightarrow \infty$. The mechanical interaction with the wall (modeled through the slip length $\ell_u$) does not appear in \cref{eq:flat_drop_velocity}, but it enters indirectly through the dependence of $\tilde{H}$ on $\tilde{\ell}_u$ (\cref{fig:TractionlessTankTreading}d): more slip yields thicker (and hence faster) droplets.

At high $\tilde{\mathcal{A}}$, the drop is flat everywhere except near the contact lines, and the local tangential traction $\tilde{\sigma}_{\text{drop/substrate}}$ induced by this drop on the substrate is identically zero almost everywhere (\cref{fig:TractionlessTankTreading}b, right panel). This is remarkable: while autonomous propulsion driven by active processes is necessarily force-free, it is not, in general, tractionless (see \cref{sec:force_traction}). Strictly speaking $\tilde{\sigma}_{\text{drop/substrate}}=0$ where $h'''=0$ (from \cref{eq:definition_traction_surface}), that is, everywhere except at the drop edges. Integrating $\tilde{\sigma}_{\text{drop/substrate}}$ over an edge yields a force of magnitude $\tilde{\phi}^2 / 2$ and directed inward: as seen from the substrate, the drop effectively acts as a contractile force dipole, independent of activity and due to finite contact angle $\tilde{\phi}$.

A sketch of the tractionless motion of a flat drop is provided in \cref{fig:TractionlessTankTreading}(e), where we also illustrate the role played by the winding number $\omega$. The winding of the director (green rods) induces an active stress in the liquid which must be balanced by the viscous stress such that the total shear stress vanishes.
The internal fluid flow thereby generated is sinusoidal, rather than parabolic in other modes of motion (blue arrows, plotted in the co-moving frame of reference). Going back to the original dimensional variables, the fluid velocity reads, in the laboratory frame of reference,
\begin{equation}
 u_x = V \left[ 1 - \cos \left( \frac{\omega \pi z}{H} \right) \right]
\end{equation}
where $H$ is the drop height and where
\begin{equation}
  V = \frac{\alpha H}{2 \pi \omega \eta}.
\end{equation}
The net flow is not zero and causes the drop to move at a velocity $V$ in a tank-treading fashion while exerting no tangential traction on the surface.
The winding number controls the number of fluid circulation cells, which is exactly equal to $|\omega|$.
The drop speed is maximum for $|\omega|=1$, which corresponds to antagonist anchoring conditions for the director at the wall and at the free surface.

The case $|\omega|=2$, while less favored energetically, generates a flow which is symmetric with respect to the drop midplane. In particular, the solution has zero fluid velocity \emph{and} zero shear stress at \emph{both} boundaries, therefore it also solves the problem of a drop squeezing through a narrow channel [right panel in \cref{fig:TractionlessTankTreading}(e)]. In this configuration the drop motion is completely independent of the amount of slip at the walls, since the drop height is geometrically constrained. This solution is reminiscent of contraction-based amoeboid motility such as exhibited by leukocyte and human breast cancer cells squeezing through complex 3D extracellular geometries \cite{Lammermann2008,Poincloux2011} and by confined cells migrating in microchannels \cite{Liu2015,Hung2013}.

This geometrically constrained setup is perhaps the easiest to control experimentally: one can imagine confining a drop of bacterial suspension~\cite{Lopez2015,Guo2018} or of microtubule-kinesin mixture~\cite{Sanchez2012} between two surfaces, one used for imaging the traction maps \cite{Dembo1996,Tan2003,duRoure2005} and the other designed to ensure appropriate anchoring (through, e.g., manipulation of the surface chemistry or architecture \cite{Koumakis2014,Sipos2015,Yuan2017,Munoz-Bonilla2018,Hasan2013}). Traction maps would show a zero traction on the channel walls everywhere except at the drop edges, where the traction magnitude and sign would only depend on the wettability of the walls.

It is important to emphasize that tractionless motion controlled by $\omega \neq 0$ is not related to the spontaneous flow transition in active nematics films \cite{Voituriez2005}. Our analysis describes a drop of active nematic in the strong elastic limit, that is, in a regime where $\mathcal{K} \geqslant O(1)$ (see Appendix \ref{app:validity_regime}), where $\mathcal{K} = (\Gamma K)/(U L)$ with $1/\Gamma$ the rotational viscosity and $K$ the nonequilibrium analog of an elastic constant. In this limit, there is no internal flow (and hence no drop motion) for $\omega=0$. It is well-known that for $\omega=0$, internal flows can occur spontaneously in thin films of active nematics beyond a critical height due to a splay (or bend) instability \cite{Voituriez2005}. Within our framework and with our notations, this instability requires $\mathcal{K} \leqslant O(\epsilon)$, in other words, it requires a drop thicker than the one we consider here. Whether the spontaneous flow transition for active films \cite{Voituriez2005} results in a ``spontaneous tractionless motion transition'' for active drops remains an open question. To answer it, one must first integrate the full dynamic equation for the director [\cref{eq:full_director_eq}] rather than its strong elastic limit [\cref{eq:director_eq}]. This problem, significantly more intricate, is left to future work.

In any case it is of fundamental importance to note that this kind of motion is \emph{only} possible for active matter driven in the bulk and cannot happen for propulsion due to driving at or near boundaries (the other two modes considered in this paper). 
 
%%%%

\subsection{Self-propulsion driven by self-advection \label{sec:AC}}

\begin{figure*}
    \flushleft
    (a) \\
    \centering
    \includegraphics[height=3.1cm]{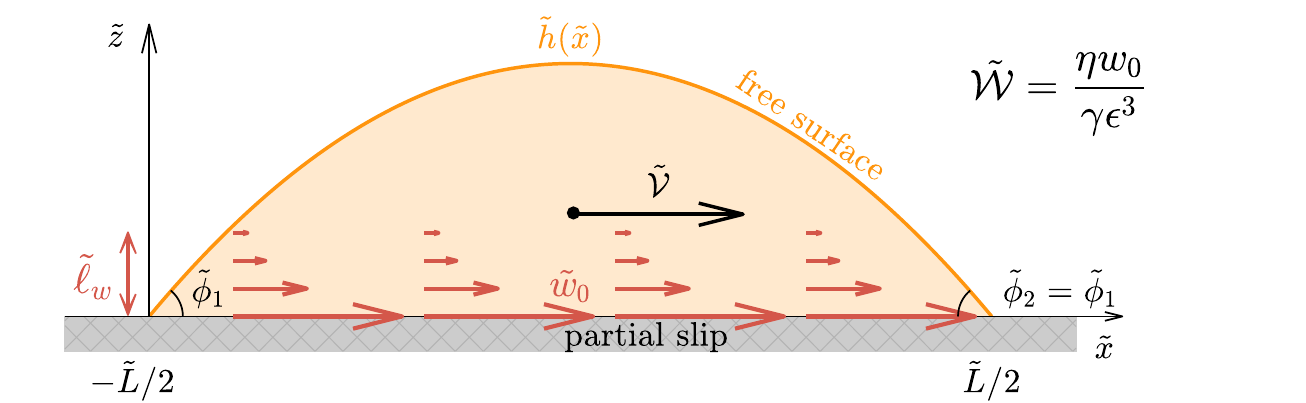} \\
    \flushleft
    (b) \\
    \centering
    \includegraphics[width=0.99\linewidth]{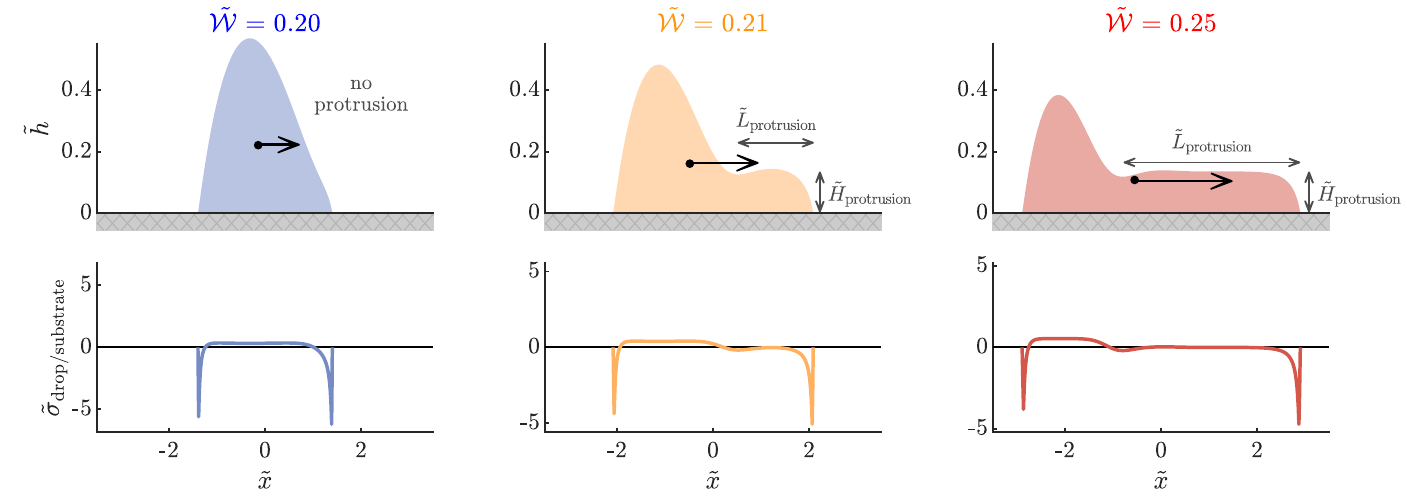} \\
    \flushleft 
    (c) \hspace{11.8cm} (d) \\
    \centering
    \includegraphics[width=0.66\linewidth]{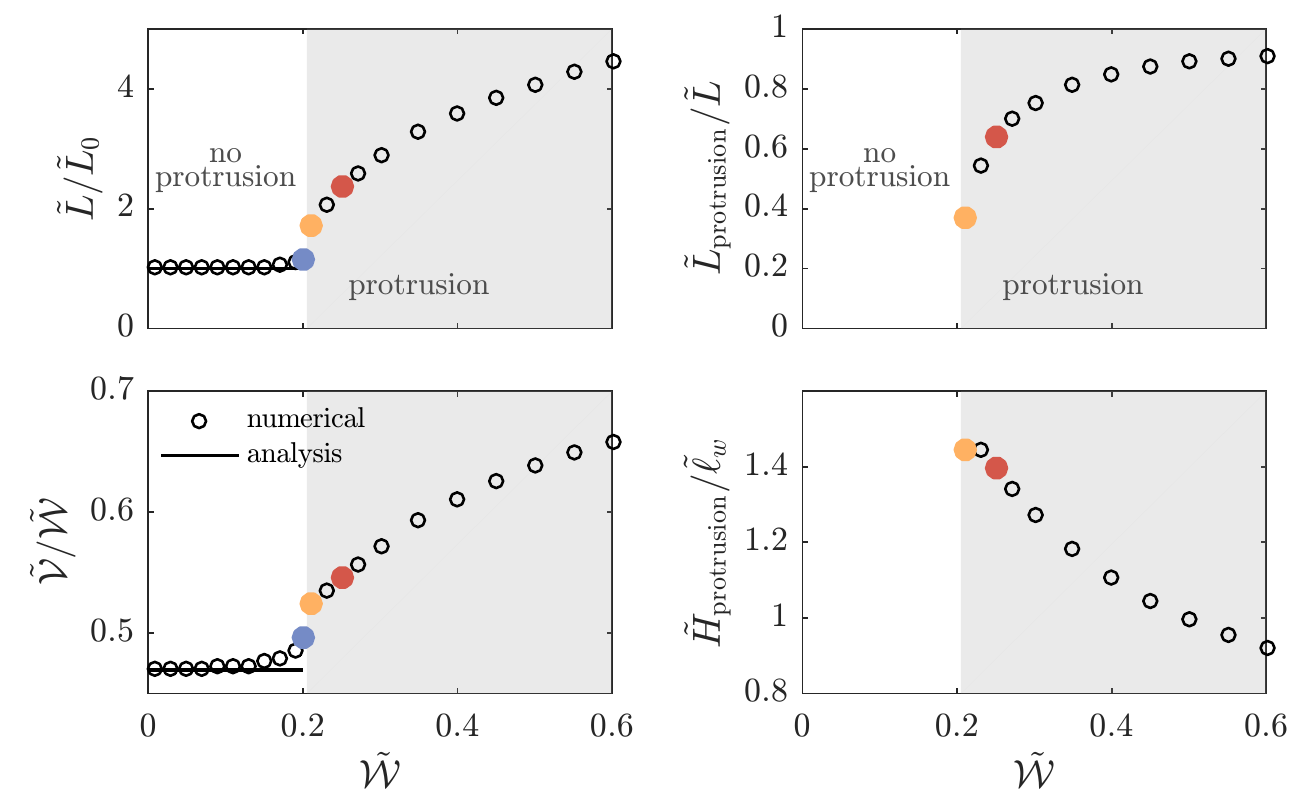} 
    \includegraphics[width=0.33\linewidth]{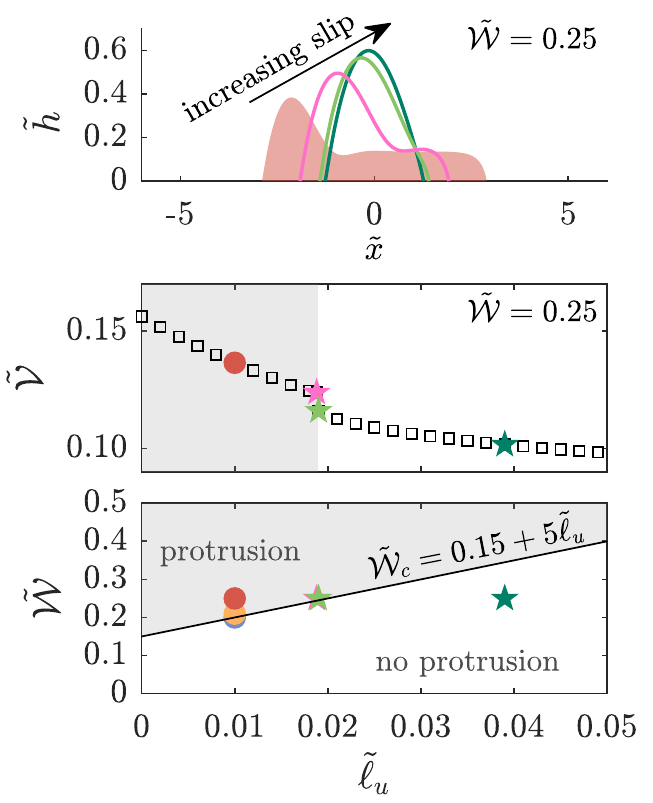} 
    \caption{ 
	Self-propulsion driven by self-advection: a drop with directed self-advection of active units (due to e.g. polymerization toward the front) develops a frontal protrusion and crawls more effectively as slip is reduced.
	(a) Model of a thin active drop with self-advection close to the substrate. 
	The effect of self-advection on the drop shape and velocity is controlled by $\tilde{\mathcal{W}}=(w_0 \eta) / (\gamma \epsilon^3)$ where $w_0$ is the characteristic self-advection speed of the active units. The height over which the strength of self-advection decays was kept constant across simulations ($\tilde{\ell}_w=0.1$). 
	(b) Numerical profiles of the drop shape and tangential traction exerted on the substrate. 
	(c) Effect of $\tilde{\mathcal{W}}$ on the drop shape and velocity, the analytical solution is given by \cref{eq:AC_pert_an}.
	(d) Effect of slip on the drop shape and velocity (top and middle panels, $\tilde{\mathcal{W}}=0.25$), and on the transition between non-protruded and protruded drops (bottom panel).
	Colored symbols in (b-d) mark corresponding state points across panels.
	\label{fig:AdhesiveCrawling}
	}
\end{figure*}

Crawling is a mode of cell motility well-characterized experimentally \cite{Verkhovsky1999,Loisel1999,Yam2007} and captured by various physical models \cite{Kruse2006,Keren2008,Shao2012,Ziebert2012,Tjhung2015,Khoromskaia2015}. % check refs
Crawling motility is usually understood as follows: polymerization of actin filaments in a thin protrusion at the leading edge generates a pushing force against the cell membrane, which, when combined with anchoring to the substrate via focal adhesions, causes the cell to move forward. 

A simple way to account for this mechanism, illustrated in \cref{fig:AdhesiveCrawling}(a), consists in adding a self-advection term to the mass conservation equation (which describes the net polymerization of filaments in a given direction), while adhesion is controlled by the amount of slip at the substrate (here through the slip length $\ell_u$). To ease comparison with prior work \cite{Tjhung2015} we chose an advection velocity which is maximum at the substrate (denoted $w_0$) and decays exponentially over a characteristic length $\ell_w$ in the direction normal to the substrate, as described in \cref{sec:self_advection}. 

We emphasize that what generates motion here is a flux of matter: crawling can be obtained solely from self-advection, in the absence of active stresses ($\tilde{\mathcal{A}}=0$) or mismatch in the contact angles ($\tilde{\phi}_1 = \tilde{\phi}_2$). The governing equation then reduces to
\begin{equation}
\label{eq:AC}
  \left( \frac{\tilde{h}}{3} + \tilde{\ell}_u \right) \tilde{h} \tilde{h}''' + \tilde{\mathcal{W}} \frac{\tilde{\ell}_w}{\tilde{h}} \left[ 1 - \exp \left( - \tilde{h}/\tilde{\ell}_w \right) \right] = \tilde{\mathcal{V}}.
\end{equation}
Besides $\tilde{\ell}_w$ (which is kept constant here, $\tilde{\ell}_w=0.1$) and $\tilde{\ell}_u$ (set here to $\tilde{\ell}_u = 0.01$), the drop shape and velocity are controlled by a single dimensionless group: $\tilde{\mathcal{W}}=(\eta w_0)/(\gamma \epsilon^3)$.

Numerical solutions of \cref{eq:AC}, computed for a range of $\tilde{\mathcal{W}}$, are presented in \cref{fig:AdhesiveCrawling}(b,c). They reveal the existence of a critical value of $\tilde{\mathcal{W}}$, denoted $\tilde{\mathcal{W}}_c$, above which a protrusion develops at the front. The numerical value of $\tilde{\mathcal{W}}_c$ increases linearly with $\tilde{\ell}_u$ [\cref{fig:AdhesiveCrawling}(d), bottom panel], and the transition is sharper for larger $\tilde{\ell}_u$, in agreement with prior work \cite{Tjhung2015}.

For $\tilde{\mathcal{W}}<\tilde{\mathcal{W}}_c$, the drop profile is nearly parabolic and it is possible to derive an analytical expression of the drop velocity at linear order in $\tilde{\mathcal{W}}$ (Appendix \ref{app:AC_perturbation_solution}):
\begin{equation}
\label{eq:AC_pert_an}
\begin{split}
 & \tilde{\mathcal{V}} = \tilde{\mathcal{W}} \frac{2 \tilde{\ell}_w}{\tilde{L}_0 \tilde{\phi}} \frac{1}{(\tilde{b}-1) \arctanh(1/\sqrt{\tilde{b}})} \Bigg\{ \\
 & \phantom{+} 2 \sqrt{\tilde{b}} \tilde{d} \,_2F_2(\{1,1\},\{3/2,2\},-\tilde{d}) \\
 & + 2 \big[ \mathrm{e}^{(\tilde{b}-1) \tilde{d}} - 1 \big] \ln\Big( 1 + 2 \big[\cot \big( \arcsin(1/\sqrt{\tilde{b}}) / 2 \big) - 1 \big]^{-1} \Big) \\
 & - \mathrm{e}^{(\tilde{b}-1) \tilde{d}} \Big[ 2 \arctanh(\sqrt{\tilde{b}}) + \mathrm{i} \pi \big[ 1 + 4 T(\sqrt{2\tilde{b}\tilde{d}},\mathrm{i}/\sqrt{\tilde{b}}) \big] \Big] \Bigg\}
\end{split}
\end{equation}
with $\mathrm{i}^2=-1$, $\tilde{b}=1 + 12 \tilde{\ell}_u/(\tilde{L}_0 \tilde{\phi})$, $d=(\tilde{\phi} \tilde{L}_0)/(4 \tilde{\ell}_w)$, and where $T(\chi,c)$ is Owen's T function \cite{Owen1956} and $_2F_2(\{a_1,a_2\},\{b_1,b_2\},\zeta)$ is the generalized hypergeometric function \cite{NIST2010book} (these functions are implemented in Mathematica). The first correction to the drop width is quadratic, so at linear order $\tilde{L}=\tilde{L}_0=\sqrt{6 \tilde{\Omega}/\tilde{\phi}}$.
Comparison to the numerical solution is shown in \cref{fig:AdhesiveCrawling}(c) (left panels): agreement is excellent nearly up to $\tilde{\mathcal{W}}_c$.

For $\tilde{\mathcal{W}}>\tilde{\mathcal{W}}_c$, the drop has a frontal protrusion of thickness $\sim \tilde{\ell}_w$ which grows in length upon increasing $\tilde{\mathcal{W}}$ [\cref{fig:AdhesiveCrawling}(c), right panels]. The drop velocity magnitude is of the order of $\tilde{\mathcal{W}}$, but its growth with $\tilde{\mathcal{W}}$ is faster than linear. 
Increasing slip reduces the drop velocity [\cref{fig:AdhesiveCrawling}(d), middle panel]: as expected crawling is most effective when the substrate provides strong adhesion.

We finally emphasize that these results, here obtained under the thin drop approximation, are in very good agreement with prior full numerical simulations \cite{Tjhung2015} (Fig.~1 and S1 therein).

\subsection{Self-propulsion driven by capillarity and modulated by activity \label{sec:CS}}

\begin{figure*}%[tbhp]
    \flushleft
    (a) \\
    \centering
     \includegraphics[height=3.1cm]{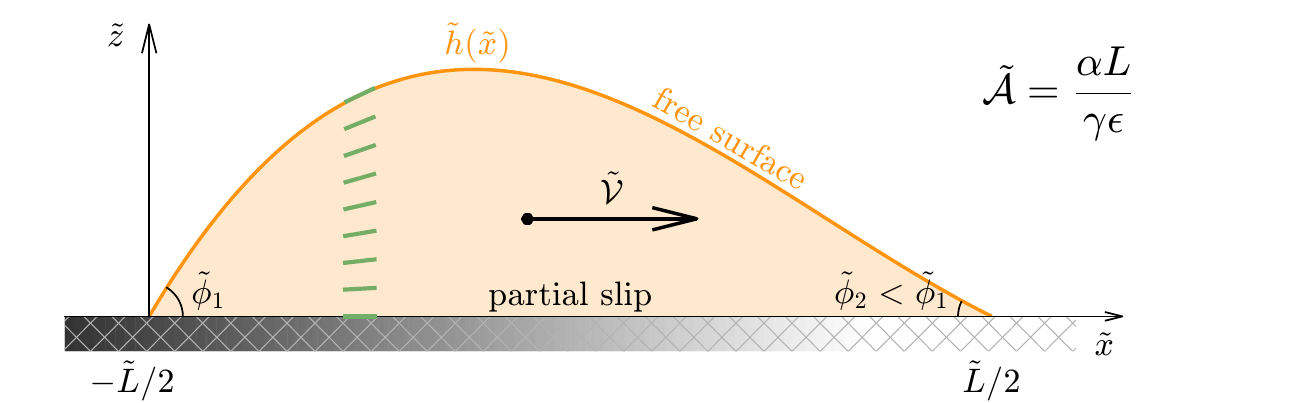}
    \includegraphics[height=3.1cm]{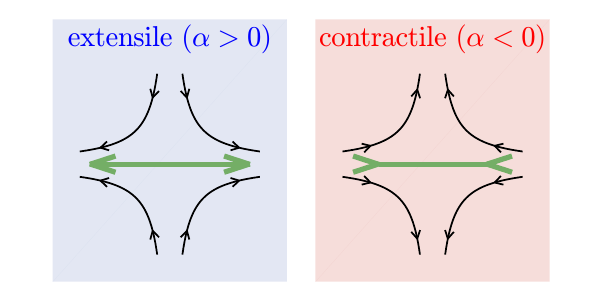}  \\
    \flushleft
    (b) \\
    \centering
    \includegraphics[width=0.99\linewidth]{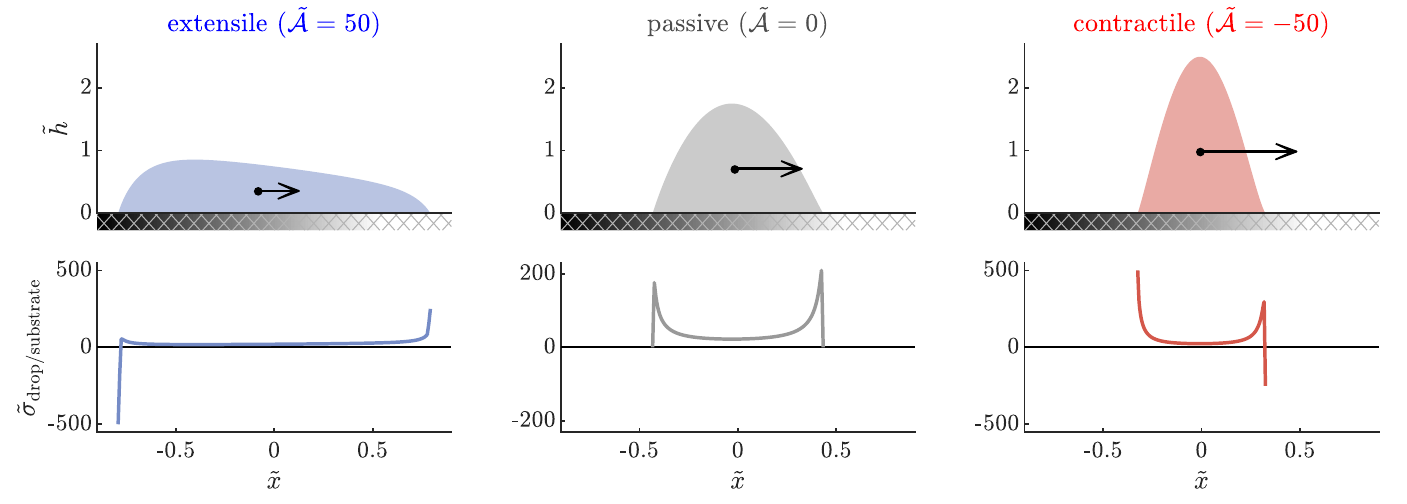} \\
    \flushleft
    (c) \hspace{11.5cm} (d) \\
    \centering
    \includegraphics[width=0.325\linewidth]{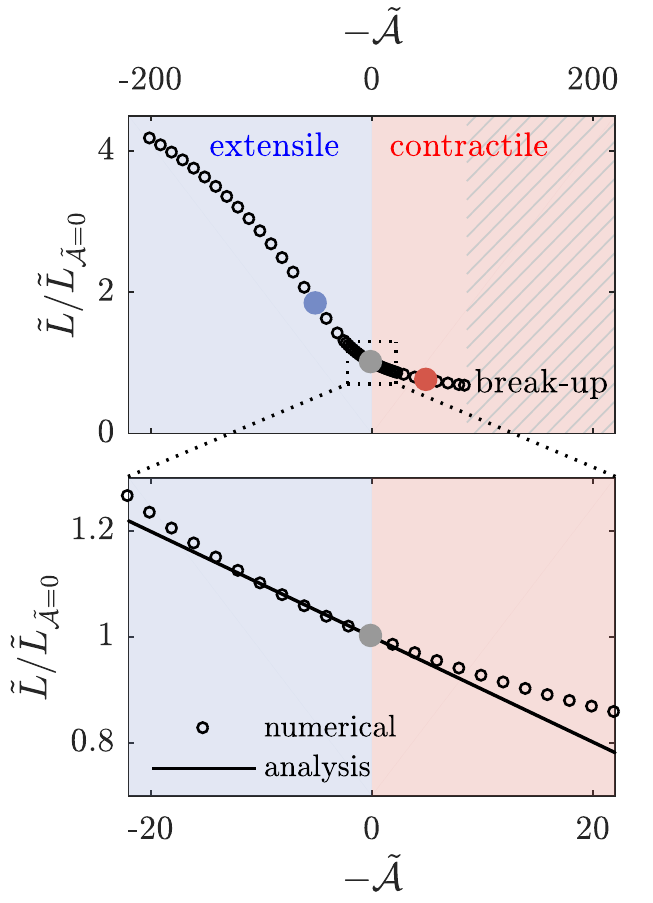} 
    \includegraphics[width=0.325\linewidth]{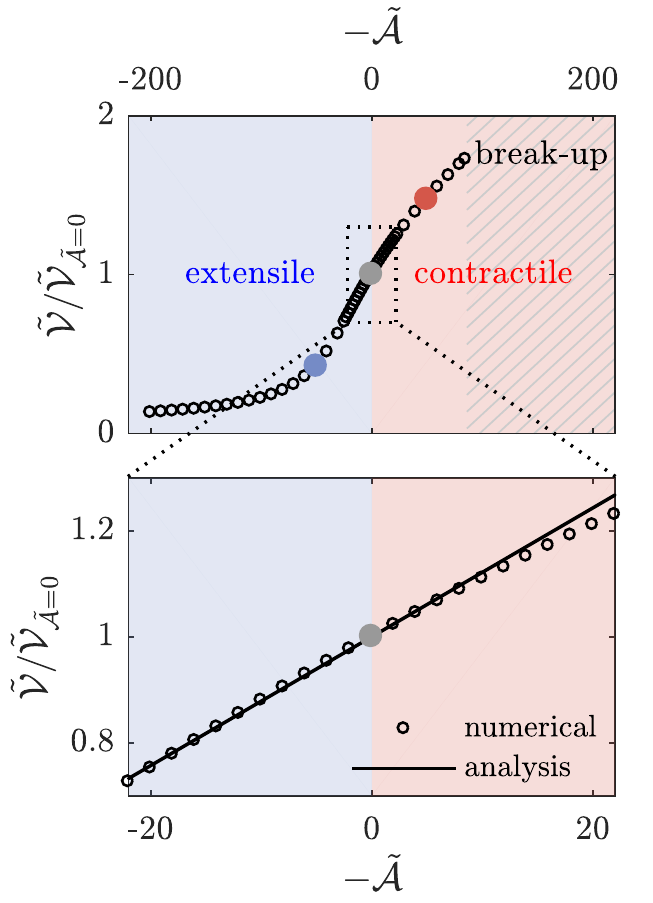} 
    \includegraphics[width=0.325\linewidth]{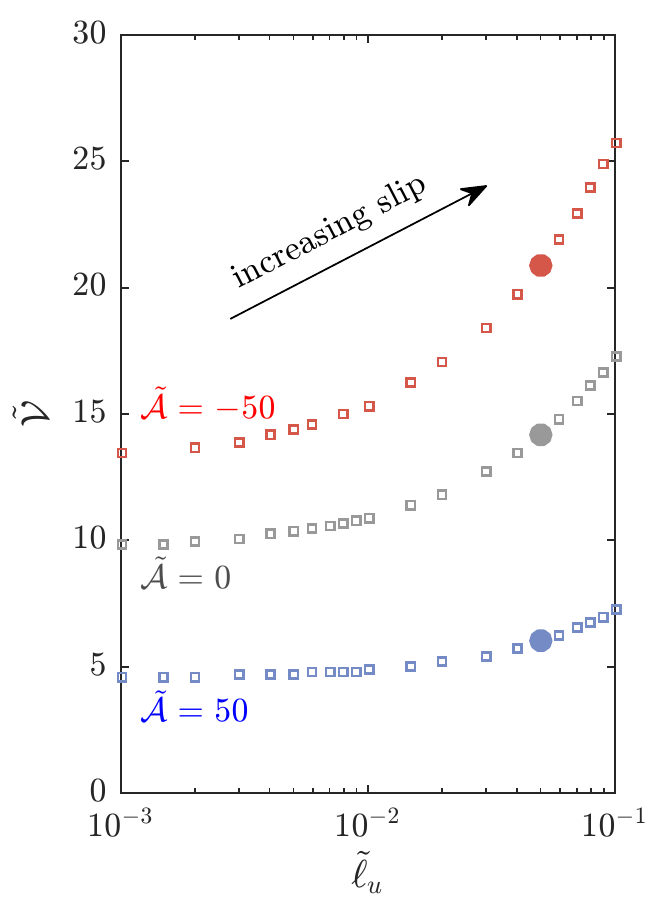}
    \caption{
    Self-propulsion driven by capillarity and modulated by activity: the motion of an asymmetric drop is enhanced by contractility and hindered by extensility.
	(a) Model of a thin active drop driven by the capillary force that results from the difference in contact angles (here $\tilde{\phi}_1=10$ and $\tilde{\phi}_2=5$) that may arise from, e.g., a gradient of surface energy. Parallel anchoring of the director is prescribed at both bounding surfaces.
	The effect of activity on the drop shape and velocity is controlled by $\tilde{\mathcal{A}}=(\alpha L) / (\gamma \epsilon)$, whose sign depends on whether the active stress is extensile ($\alpha > 0$) or contractile ($\alpha < 0$). 
	(b) Numerical profiles of the drop shape and tangential traction exerted on the substrate.
	(c) Effect of $\tilde{\mathcal{A}}$ on the drop width and velocity with respect to the passive case, analytical expressions are given by \cref{eq:CS_pert_an_L} and \cref{eq:CS_pert_an_V}. 
	(d) Effect of slip on the drop velocity.
	Colored symbols in (b-d) mark corresponding state points across panels.
	\label{fig:CapillarySliding} 
	}
\end{figure*}

We consider an asymmetric drop moving under the action of a net capillary force ($\phi_{1} \neq \phi_{2}$ implies $\abs{F_{\text{capillary}}}>0$), as depicted in \cref{fig:CapillarySliding}(a). The director field is chosen to be nearly aligned ($\omega=0$), as larger distortions would essentially lead back to \cref{sec:TTT} (motion driven by active stresses and controlled by the winding of the director). Here we ask: can activity facilitate (or impede) the drop motion, and does the sign of activity (extensile or contractile) matter? The answer: Yes, and yes.

A minimal mathematical description of capillarity-driven sliding is obtained by setting $\tilde{\mathcal{W}}=0$ and $\omega=0$ in \cref{eq:scaled_ODE}, which yields
\begin{subequations}
\label{eq:CS-A}
\begin{equation}
    \left( \frac{\tilde{h}}{3} + \tilde{\ell}_u \right) \tilde{h} \tilde{h}''' - \tilde{\mathcal{A}} \left( \frac{(3-m)\tilde{h}}{6} + \tilde{\ell}_u \right) \tilde{h}' = \tilde{\mathcal{V}} 
\end{equation}
with different contact angles imposed at the boundaries (as may arise, from, e.g., a gradient of surface energy):
\begin{equation}
 \tilde{\phi}_1 = \tilde{\phi} + \frac{\tilde{\varphi}}{2},  \qquad \tilde{\phi}_2 = \tilde{\phi} - \frac{\tilde{\varphi}}{2}
\end{equation}
\end{subequations}
with $\tilde{\varphi}=\tilde{\phi}_1-\tilde{\phi}_2$ the contact angle difference, as depicted in \cref{fig:CapillarySliding}(a).
The effect of activity on the drop shape and velocity is controlled by the dimensionless parameter $\tilde{\mathcal{A}} = (\alpha L)/(\gamma \epsilon)$.
The sign of $\tilde{\mathcal{A}}$ depends on whether active units, modelled as force dipoles \cite{Pedley1992}, induce an extensile flow ($\alpha > 0$, e.g. certain bacteria) or a contractile flow ($\alpha < 0$, e.g. the actin-myosin complex).

Numerical solutions to \cref{eq:CS-A} for various $\tilde{\mathcal{A}}$ are presented in \cref{fig:CapillarySliding}(b,c). With respect to the passive case, the drop base is narrower (wider) with contractile (extensile) activity [\cref{fig:CapillarySliding}(b,c)], as is the case for static symmetric drops \cite{Joanny2012}. At high $-\tilde{\mathcal{A}}$ (high contractility), the drop breaks up; this phenomenon is outside the scope of this paper and its analysis is left to future work. 
Activity also influences the drop speed: contractile (extensile) drops are faster (slower) than their passive counterpart [\cref{fig:CapillarySliding}(c)] (note that with normal anchoring, the extensile drop would be narrower and faster). Increasing slip results in greater drop velocities, as for a passive drop since friction hinders sliding motion [\cref{fig:CapillarySliding}(d)].

The dependence of $\mathcal{\tilde{V}}$ and $\tilde{L}$ on $\tilde{\mathcal{A}}$ is approximately linear over a rather large range of $\tilde{\mathcal{A}}$, and can be computed exactly from a perturbation analysis in the limit of small $|\tilde{\varphi}|$ and small $|\tilde{\mathcal{A}}|$. 
We find (Appendix \ref{app:CS_perturbation_solution}) that the drop width is
\begin{subequations}
\label{eq:CS_pert_an_L}
\begin{equation}
  \tilde{L} = \tilde{L}_0 + \tilde{\mathcal{A}} \tilde{L}_\alpha + O(\tilde{\mathcal{A}}^2, \tilde{\varphi}^2)
\end{equation}
with ($\Omega$ is the drop volume)
\begin{equation}
  \tilde{L}_0 = \sqrt{\frac{6 \tilde{\Omega}}{\tilde{\phi}}}, \qquad \tilde{L}_\alpha = \frac{\tilde{\Omega}}{2 \tilde{\phi}^2}
\end{equation}
\end{subequations}
and the drop velocity is
\begin{subequations}
\label{eq:CS_pert_an_V}
\begin{equation}
  \tilde{\mathcal{V}} = \tilde{\varphi} \left[ \tilde{\mathcal{V}}_{\varphi,L_0} + \tilde{\mathcal{A}} (\tilde{\mathcal{V}}_{\varphi,L_\alpha}  + \tilde{\mathcal{V}}_{\varphi,\alpha}) \right] + O(\tilde{\mathcal{A}}^2, \tilde{\varphi}^2)
\end{equation}
where the coefficients are given below:
\begin{align}
	\tilde{\mathcal{V}}_{\varphi,L_0} & = \frac{\tilde{\beta}_0 \tilde{\phi}}{6 \ln (\tilde{\beta}_{p/m})} \\[0.8em]
	\tilde{\mathcal{V}}_{\varphi,L_\alpha} & = - \frac{\tilde{\phi}^2 \tilde{L}_\alpha [ \tilde{\beta}_0 \tilde{L}_0 + 6 \tilde{\ell}_u \ln (\tilde{\beta}_{p/m}) ] }{ 6 \tilde{L}_0^2 \tilde{\beta}_0 [\ln (\tilde{\beta}_{p/m})]^2}\\[0.8em]
	\tilde{\mathcal{V}}_{\varphi,\alpha} & = \frac{\tilde{L}_0 \tilde{\beta}_{p/m}}{24 \tilde{\beta}_0 [ \ln (\tilde{\beta}_{p/m}) ]^2} \Bigg\{ \\
	& + ( 2\tilde{\beta}_0^2 - \tilde{\beta}_{m}^2 ) \dilog ( \tilde{\beta}_{p/m} ) \nonumber \\
	& - ( 2\tilde{\beta}_0^2 + \tilde{\beta}_{m}^2 ) \dilog ( 2 \tilde{\phi}/\tilde{\beta}_{p} ) \nonumber \\
	& - \tilde{\beta}_0^2 \big\{ [ \ln(\tilde{\beta}_{p}) ]^2 - [ \ln(\tilde{\beta}_{m}) ]^2 \big\} \nonumber \\
	& + \big\{ [ 2 \tilde{\beta}_0^2 + \tilde{\beta}_{m}^2 ] \ln(2\tilde{\phi}) - \tilde{\beta}_{m}^2 [ 1 + \ln(\tilde{\beta}_{p}) ] \big\} \ln (\tilde{\beta}_{p/m}) \nonumber \\
	& + \frac{\pi^2}{6} ( 2 \tilde{\beta}_0^2 + \tilde{\beta}_{m}^2 ) - 2 \frac{\tilde{\beta}_0 \tilde{\phi}}{\tilde{\beta}_{p/m}} \Bigg\} \nonumber 
\end{align}
\end{subequations}
with $\tilde{\beta}_0 = \tilde{\phi} \sqrt{1+(12 \tilde{\ell}_u)/(\tilde{L}_0 \tilde{\phi})}$, $\tilde{\beta}_{p} = \tilde{\beta}_0 + \tilde{\phi}$, $\tilde{\beta}_{m} = \tilde{\beta}_0 - \tilde{\phi}$, $\tilde{\beta}_{p/m}= \tilde{\beta}_{p}/\tilde{\beta}_{m}$ and with $\dilog (y) = \displaystyle \int_1^y \dfrac{\ln(t)}{1-t} \id t$.
Comparison with the numerical solution is excellent [bottom panels in \cref{fig:CapillarySliding}(c)].

Embedding an active suspension into an otherwise self-propelled passive droplet, driven by a gradient of surface energy or surface tension \cite{Chaudhury1992,Bain1994,DosSantos1995,Cira2015}, could be a rather straightforward way to realize experimentally this activity-modulated, capillarity-driven self-propulsion.

\section{Conclusions}

To summarize, we have obtained a generic and unified description of a thin 2D drop of active liquid moving on a solid substrate that consists of a single ODE. We have analyzed, using a combination of numerical simulations and asymptotic analysis, the autonomous propulsion of this drop induced by three possible driving sources (summarized in \cref{fig:scheme_intro}): active stresses (\cref{fig:TractionlessTankTreading}), active self-advection (\cref{fig:AdhesiveCrawling}), and a (possibly self-generated) capillary force (\cref{fig:CapillarySliding}).

Motion driven by active stresses does not require a shape asymmetry, is efficient even in the presence of slip and allows self-propulsion without the need to exert traction anywhere on the surface, giving rise to ``tractionless tank treading''. This new mode of motion, driven in the bulk rather than at the boundaries, is topologically protected and is particularly suited for moving rapidly through tiny pores. Therefore it provides a robust physical mechanism for efficient cell migration in tissues.

In contrast, motion driven by the self-advection of polarized active units at the substrate, known as crawling, is characterized by a strong shape anisotropy and is most efficient in the absence of slip. Therefore this mode of self-propulsion is particularly suited for moving on 2D surfaces which provides strong anchoring points. A prominent example of crawling is mesenchymal migration, a mode of cell motility characterized by strong cell-substrate adhesions, with self-advection provided by actin polymerization in a leading edge protrusion \cite{Verkhovsky1999,Loisel1999}.

Finally the third mode motion is, unlike the other two, not driven by active processes but by a net capillary force, as can be induced by (external or self-generated) thermal or chemical gradients. By coupling this driving with internal activity, one can further tune the drop velocity and create droplets faster than their passive counterparts.

Our 2D model is expected to be valid for any 3D drop where variations in the additional spatial dimension are much slower than in the other two. Beyond that, while extending the thin drop formulation to 3D is rather straightforward, solutions may be far more complex (for example, based on the form of \cref{eq:TTT}, we expect a fingering instability for a 3D drop driven by active stresses).

Our present attempt to provide a generic classification of self-propulsion mechanisms, one of them being the extensively studied treadmilling-driven crawling, led us to introduce an entirely new class (tractionless tank-treading driven by bulk active stresses) and to give a new twist to an old mechanism (self-propulsion driven by gradients): we hope those will trigger further theoretical investigations and experimental realizations.

\appendix

\section{Unsteady height evolution \label{app:height_equation}} 

Consider the general (unsteady) case of a drop shape described by a height function $h^*(x^*,t)$ in the laboratory frame of reference (denoted by a star). It is related to the flow at the free surface by a kinematic boundary condition \cite{Oron1997,Eggers2015book}
\begin{equation}
	\partial_t h^* + (u_x + w n_x) \partial_{x^*} h^* - (u_z + w n_z) = 0
	\label{eq:kin_BC}
\end{equation}
with $\v{u}$ the fluid velocity and $w \v{n}$ the self-advection at speed $w$ of active units with local orientations $\v{n} = (\cos \theta, \sin \theta)$.
At the free surface, the anchoring angle is $\theta = \omega \frac{\pi}{2} + \arctan(\partial_{x^*} h^*)$.
It follows that
\begin{equation}
	n_x \partial_{x^*} h^* - n_z = - \sin \left( \frac{\omega \pi}{2} \right) \sqrt{1+(\partial_{x^*} h^*)^2}.
\end{equation}
Besides, using flow incompressibility and wall impermeability, one can show that
\begin{equation}
	u_x \partial_{x^*} h^* - u_z = \partial_{x^*} \int_0^{h^*} u_x \; \d z.
\end{equation}
Therefore the kinematic boundary condition \cref{eq:kin_BC} can be rewritten as
\begin{equation}
	\partial_t h^* + \partial_{x^*} \int_0^{h^*} u_x \; \d z = w (h^*) \sin \left( \frac{\omega \pi}{2} \right) \sqrt{1+(\partial_{x^*} h^*)^2}
	\label{eq:height_eq_PDE}
\end{equation}
In this paper we restrict to either $w=0$ or $\omega=0$, so the right-hand-side of \cref{eq:height_eq_PDE} is zero.

\section{Regime of validity \label{app:validity_regime}}

The evolution of the director field in a nematic liquid crystal is governed by
\begin{multline}
	\partial_t n_i + [ (u_j + w n_j) \partial_j ]n_i - \Omega_{ij} n_j = \\
	\delta_{ij}^T ( \lambda E_{jk} n_k + \Gamma K \nabla^2 n_j )
	\label{eq:full_director_eq}
\end{multline}
where $\delta_{ij}^T=\delta_{ij}-n_i n_j$ is a transverse projection operator, $E_{ij}=(\partial_i u_j + \partial_j u_i)/2$ and $\Omega_{ij}=(\partial_j u_i - \partial_i u_j)/2$ are the strain-rate and rotation-rate tensors, $\lambda$ is the flow alignment parameter, $1/\Gamma$ is the rotational viscosity and $K$ is the nonequilibrium analog of a Frank constant. 

Momentum conservation reads, including inertial terms,
\begin{equation}
   \rho (\partial_t + u_j \partial_j) u_i = \partial_j \sigma_{ij}
\end{equation}
where $\rho$ is the mass density of the fluid and where $\sigma_{ij}$ is given by \cref{eq:stress_definition} with $\sigma_{ij}^n = - K \partial_i n_k \partial_j n_k$.

In addition to $\epsilon$ and $\phi_{1,2}$, the dynamics of the drop is controlled by seven dimensionless groups that can be constructed from those equations: $\lambda$, $\mathcal{W} = w_0/U$, $\mathcal{K} = (\Gamma K)/(U L)$, $\mathcal{A} = (\alpha L)/(\eta U)$, $\mathcal{C} = \gamma/(\eta U)$, $\mathcal{N} = K/(\eta U L)$ and $\mathcal{R} = \rho U L/\eta$. The equations we solve here, presented in \cref{sec:model}, are obtained when the dimensionless groups satisfy the conditions summarized in \cref{tab:validity_regime}. They describe a thin drop whose dynamics is determined from the balance of active, viscous and surface tension forces in a regime where inertia and nematic stresses are negligible and where the director field is not back-coupled to the flow.

\begin{table}[!h]
\begin{tabular*}{0.48\textwidth}{@{\extracolsep{\fill}}llll}
\hline
			& \multicolumn{3}{c}{self-propulsion driven by} \\
                        & active stresses & self-advection & capillarity \\
					  & $\omega \neq 0$ & $\omega = 0$ & $\omega = 0$ \\
                                            & $\theta=O(1)$ & $\theta=O(\epsilon)$ & $\theta=O(\epsilon)$ \\
  \hline
  $\mathcal{W} = w_0/U$                                 & $\leqslant O(\epsilon)$ & $=O(1)$ & $\leqslant O(\epsilon)$ \\               
  $\mathcal{A} = (\alpha L)/(\eta U)$   & $=O(\epsilon^{-1})$ & $\leqslant O(\epsilon^{-1})$ & $=O(\epsilon^{-2})$ \\
  $\mathcal{C} = \gamma/(\eta U)$       & $=O(\epsilon^{-3})$ & $=O(\epsilon^{-3})$ & $=O(\epsilon^{-3})$ \\
  $\mathcal{N} = K/(\eta U L)$          & $\leqslant O(\epsilon)$ & $\leqslant O(\epsilon^{-1})$ & $\leqslant O(\epsilon^{-1})$ \\
  $\mathcal{R} = \rho U L/\eta$         & $\leqslant O(\epsilon^{-1})$ & $\leqslant O(\epsilon^{-1})$ & $\leqslant O(\epsilon^{-1})$ \\
  $\mathcal{K} = (\Gamma K)/(U L)$      & $\geqslant O(1)$ & $\geqslant O(\epsilon^{-1})$ & $\geqslant O(\epsilon^{-1})$ \\
  $\lambda$                             &  $\leqslant O(1)$ & $\leqslant O(1)$ & $\leqslant O(1)$ \\
  \hline
\end{tabular*}
\caption{Range of validity of our analysis in terms of the dimensionless groups governing the drop dynamics. \label{tab:validity_regime}}
\end{table}

\section{Force balance on the drop \label{app:force_balance}}

The sum of the forces exerted on the drop, denoted $\v{F}_{\text{total}}$, satisfies $\v{F}_{\text{total}}= \v{0}$ (as follows from integrating \cref{eq:momentum_conservation} over the drop) and is defined by
\begin{subequations}
\begin{equation}
	\v{F}_{\text{total}} = \v{F}_{\text{substrate/drop}} + \v{F}_{\text{free surface/drop}}
\end{equation}
with
\begin{align}
 \v{F}_{\text{substrate/drop}} & = \int_{\partial \mathcal{D}_\text{sol/liq}} \m{\sigma} \bcdot \m{m} \id s \\
 \v{F}_{\text{free surface/drop}} & = \int_{\partial \mathcal{D}_\text{gas/liq}} \m{\sigma} \bcdot \m{m} \id s
\end{align}
\end{subequations}
where $\mathcal{D}$ is the domain occupied by the drop, $\partial \mathcal{D}$ is its boundary (decomposed into solid/liquid and gas/liquid interfaces), and $\v{m}$ is the unit normal directed outward the boundary. 

The x-component of the force exerted by the substrate on the drop is
\begin{equation}
\begin{split}
    F_{\text{substrate/drop},x} & = \int_{\partial \mathcal{D}_\text{sol/liq}} - \sigma_{xz} \rvert_{z=0} \id x \\
    & = \int_{\partial \mathcal{D}_\text{sol/liq}} - \eta ( \partial_z u_x + \partial_x u_z ) \id x \\
    & \equiv F_{\text{friction}}
\end{split}
\end{equation}
Since the active contribution $\sigma_{xz}^a \rvert_{z=0}$ vanishes for parallel or normal anchoring of the director, $F_{\text{substrate/drop},x}$ is purely frictional and we denote it $F_{\text{friction}}$ in the main text.

The force exerted by the free surface on the drop is, using \cref{eq:BC_surface_tension}, 
\begin{equation}
    \v{F}_{\text{free surface/drop}} = \int_{\partial \mathcal{D}_\text{gas/liq}} \gamma \kappa \m{m} \id s
\end{equation}
and in 2D we have
\begin{equation}
 \begin{split}
	\v{m} & = \frac{1}{\sqrt{1+h'^2}} (-h',1) \\
	\kappa & = \frac{h''}{\left[ 1 + h'^2 \right]^{3/2}} \\
	\d s & = \sqrt{1+h'^2} \d x
\end{split}
\end{equation}
It is straightforward to show that in the x-direction,
\begin{equation}
\begin{split}
	F_{\text{free surface/drop},x} & = \gamma \left( \cos \phi_2 - \cos \phi_1 \right) \\
	& \equiv F_{\text{capillary}}
\end{split}
\end{equation}
where $\phi_1$ and $\phi_2$ are the contact angles on each side of the drop.
Since $F_{\text{free surface/drop},x}$ originates purely from surface tension, we refer to it as $F_{\text{capillary}}$ in the main text.

\section{Perturbation analysis}

In this section we will derive, using a perturbation analysis, the first effect of activity $\tilde{\mathcal{A}}$ or of self-advection $\tilde{\mathcal{W}}$ on the drop shape $\tilde{h}$, velocity $\tilde{\mathcal{V}}$, and width $\tilde{L}$.

To facilitate the derivation we first rescale the x-coordinates by introducing the change of variable $\tilde{y}= \tilde{\xi} \tilde{x}$ with $\tilde{\xi}=2/\tilde{L}$, and we set $m=1$.
The governing ODE for $\tilde{g}(\tilde{y})=\tilde{h}(\tilde{x})$ is then
\begin{subequations}
\label{eq:full_scaled_problem_g}
\begin{equation}
\label{eq:scaled_ODE_g}
\begin{split}
  & \tilde{\xi}^3\left( \frac{\tilde{g}}{3} + \tilde{\ell}_u \right) \tilde{g} \tilde{g}''' + \tilde{\mathcal{A}} \tilde{f}^{\alpha}(\tilde{g}) + \tilde{\mathcal{W}} \tilde{f}^w(\tilde{g}) = \tilde{\mathcal{V}} \\
% \intertext{with}
  \tilde{f}^{\alpha}(\tilde{g}) & = 
 \begin{dcases}
 \tilde{g} & \text{if $\omega \neq 0$,}\\
 - \tilde{\xi} \left( \frac{\tilde{g}}{3} + \tilde{\ell}_u \right) \tilde{g}'  & \text{if $\omega = 0$,}
\end{dcases} \\
  \tilde{f}^w(\tilde{g}) & = 
  \begin{dcases}
  	\text{not considered} & \text{if $\omega \neq 0$,}\\
  	\frac{\tilde{\ell}_w}{\tilde{g}} \left[ 1 - \exp \left( - \tilde{g}/\tilde{\ell}_w \right) \right] & \text{if $\omega = 0$,}
  \end{dcases}
\end{split}
\end{equation}
with boundary conditions
\begin{equation}
\label{eq:scaled_BC_g}
\begin{gathered}
 \tilde{g}(\pm 1) = 0, \quad \tilde{\xi} \tilde{g}'(\pm 1) = \mp \tilde{\phi} + \frac{\tilde{\varphi}}{2},
\end{gathered}
\end{equation}
where $\tilde{\varphi}=\tilde{\phi}_1-\tilde{\phi}_2$ is the difference between contact angles on each side of the drop, and with the volume constraint
\begin{equation}
 \int_{-1}^{1} \tilde{g} \; \d \tilde{x} = \tilde{\xi} \tilde{\Omega}.
 \label{eq:scaled_volume_constraint_g}
\end{equation}
\end{subequations}
In the following all quantities are scaled in the lubrication framework and we drop the tilde in the remaining of this section.

We start from the exact analytical solution, denoted by subscript $0$, for a symmetric passive drop ($\varphi=0$, $\mathcal{A}=0$, $\mathcal{W}=0$):
\begin{subequations}
\begin{align}
	g_0 & = \phi (1-y^2)/(2 \xi) \\
	L_0 & = \sqrt{6 \Omega/\phi} \\
	\mathcal{V}_0 & = 0.
\end{align}
\end{subequations}
We will then expand the solution to \cref{eq:full_scaled_problem_g} as a perturbation power series of the relevant parameters for each mode of motion:
\begin{subequations}
\begin{align}
	g & = g_0 + \delta g_\delta + O(\delta^2) \\
	L & = L_0 + \delta L_\delta + O(\delta^2) \\
	\mathcal{V} & = \mathcal{V}_0 + \delta \mathcal{V}_\delta + O(\delta^2),
\end{align}
\end{subequations}
where $\delta$ is the small parameter in which we expand.

As we shall see later on, substituting this expansion into the governing ODE and matching terms at order $\delta$ yield an ODE of the form
\begin{equation}
    \label{appeq:generic_pert_ODE}
	g_\delta'''(y) = \mathcal{V}_{\delta} R_1(y) + R_2(y)
\end{equation}
with boundary conditions
\begin{equation*}
	g_\delta(\pm 1) = 0 \qquad g_\delta'(\pm 1)= c
\end{equation*}
where $c$ is a constant which depends on the mode of motion considered.

Conveniently $\mathcal{V}_{\delta}$ and $L_\delta$ can be computed without solving \cref{appeq:generic_pert_ODE} by using a solvability condition in the spirit of the approach presented in \cite{Pismen2008}.
We introduce a test function $t(y)$ which satisfies $t(\pm 1)=0$. We can write
\begin{equation}
 \int_{-1}^{1} t g_\delta''' \id y = \mathcal{V}_{\delta} \int_{-1}^{1} t R_1 \id y + \int_{-1}^{1} t R_2 \id y
\end{equation}
The left-hand-side can be integrated by part three times to yield
\begin{equation}
 \int_{-1}^{1} t g_\delta''' \id y = - c \big[ t' \big]_{-1}^{1}  - \int_{-1}^{1} t''' g_\delta \id y
\end{equation}
Choosing an adequate test function which satisfies $t'''=0$, we can determine $\mathcal{V}_\delta$ from
\begin{equation}
  - c \big[ t' \big]_{-1}^{1} = \mathcal{V}_{\delta} \int_{-1}^{1} t R_1 \id y + \int_{-1}^{1} t R_2 \id y.
\end{equation}
On the other hand, choosing a test function with $t'''=\delta$ yields 
\begin{equation}
 - c \big[ t' \big]_{-1}^{1} - \delta \int_{-1}^{1} g_\delta \id y = \mathcal{V}_{\delta} \int_{-1}^{1} t R_1 \id y + \int_{-1}^{1} t R_2 \id y
\end{equation}
From the volume constraint one has
\begin{equation}
	 \delta \int_{-1}^{1} g_\delta \id y = \xi \Omega - \int_{-1}^{1} g_0 \id y
\end{equation}
therefore we can determine $\xi$ (and then $L=2/\xi$) from
\begin{equation}
  - c \big[ t' \big]_{-1}^{1} - \xi \Omega + \int_{-1}^{1} g_0 \id y = \mathcal{V}_{\delta} \int_{-1}^{1} t R_1 \id y + \int_{-1}^{1} t R_2 \id y
\end{equation}

\subsection{Self-propulsion driven by active stresses \label{app:TTT_perturbation_solution}}

We consider here the case of motion solely driven by active stresses ($\omega \neq 0$, $\varphi=0$, $\mathcal{A} \neq 0$, $\mathcal{W}=0$), so the ODE reduces to
\begin{equation}
\label{eq:TTT_g}
  \xi^3 \left( \frac{g}{3} + \ell_u \right) g g''' + \mathcal{A} g = \mathcal{V}.
\end{equation}
We write the perturbation solution as
\begin{subequations}
\begin{align}
	g & = g_0 + \mathcal{A} g_\alpha + O(\mathcal{A}^2) \\
	L & = L_0 + \mathcal{A} L_\alpha + O(\mathcal{A}^2) \\
	\mathcal{V} & = \mathcal{V}_0 + \mathcal{A} \mathcal{V}_\alpha + O(\mathcal{A}^2).
\end{align}
\end{subequations}
At linear order the correction is solution of
\begin{subequations}
\begin{gather}
   g_\alpha''' = \frac{\mathcal{V}_\alpha - g_0}{\xi^3 \left( \dfrac{g_0}{3} + \ell_u \right) g_0} \\
   \text{with\ } \quad g_\alpha(\pm 1) = 0, \quad g_\alpha'(\pm 1) = 0,
\end{gather}
\end{subequations}
and we find after integration
\begin{subequations}
\begin{align}
	g_\alpha & =  \frac{3}{2 \phi \xi^2 (b-1) \arctanh(1/\sqrt{b}) } \bigg\{  \\
	& c \left[ (1+y)^2 \ln(1+y) - (1-y)^2 \ln(1-y) - (2^2 \ln 2) y \right] \nonumber \\
	& - \left[ (\sqrt{b}+y)^2 \ln(\sqrt{b}+y) - (\sqrt{b}-y)^2 \ln(\sqrt{b}-y) \right] \nonumber \\
	& + \left[ (\sqrt{b}+1)^2 \ln(\sqrt{b}+1) - (\sqrt{b}-1)^2 \ln(\sqrt{b}-1) \right] y  \bigg\} \nonumber \\
	L_\alpha & = 0 \\
	V_\alpha & = \frac{\phi}{2 \xi} \frac{c}{\arctanh(1/\sqrt{b})}
\end{align}
\end{subequations}
where $b=1 + 6 \ell_u \xi/\phi$ and $c=\sqrt{b} - (b-1) \arctanh(1/\sqrt{b})$.

\subsection{Self-propulsion driven by self-advection \label{app:AC_perturbation_solution}}

We consider here the case of pure advective crawling ($\omega = 0$, $\varphi=0$, $\mathcal{A}=0$, $\mathcal{W} \neq 0$), so the ODE reduces to
\begin{equation}
  \xi^3\left( \frac{g}{3} + \ell_u \right) g g''' + \mathcal{W} \frac{\ell_w}{g} \left[ 1 - \exp \left( - g/\ell_w \right) \right] = \mathcal{V}
\end{equation}
Writing the perturbative solution as
\begin{subequations}
\begin{align}
	g & = g_0 + \mathcal{W} g_w + O(\mathcal{W}^2) \\
	L & = L_0 + \mathcal{W} L_w + O(\mathcal{W}^2) \\
	\mathcal{V} & = \mathcal{V}_0 + \mathcal{W} \mathcal{V}_w + O(\mathcal{W}^2),
\end{align}
\end{subequations}
we have the following problem at linear order
\begin{subequations}
\begin{gather}
   g_w''' = \frac{\mathcal{V}_w - \dfrac{\ell_w}{g_0} \left[ 1 - \exp \left( - g_0/\ell_w \right) \right]}{\xi^3\left( \dfrac{g_0}{3} + \ell_u \right) g_0 } \\
   \text{with\ } \quad g_w(\pm 1) = 0, \quad g_w'(\pm 1) = 0.
\end{gather}
\end{subequations}
The constant $\mathcal{V}_w$ can be determined from a solvability condition. Using the test function $t=1-y^2$ we obtain
\begin{equation}
\label{eq:complicated_integral_velocity}
	\mathcal{V}_w = \frac{\xi \ell_w}{\phi} \frac{\sqrt{b}}{\arctanh(1/\sqrt{b})} I
\end{equation}
where $I$ is the definite integral
\begin{equation}
	I = \int_{-1}^{1} \frac{\{ 1 - \exp [ -d (1-y^2) ] \}}{ (1-y^2) (b-y^2)} \id y
\end{equation}
with $b=1 + 6 \ell_u \xi/\phi$ and $d=\phi/(2 \xi \ell_w)$.
A series of manipulations allowed us to obtain an explicit expression of $I$, which reads
\begin{equation}
\label{eq:nasty_integral}
\begin{split}
 & I = \frac{1}{(b-1) \sqrt{b}} \Bigg\{ 2 \sqrt{b} d\,_2F_2(\{1,1\},\{3/2,2\},-d) \\
 & + 2 \big[ \mathrm{e}^{(b-1)d} - 1 \big] \ln\Big( 1 + 2 \big[\cot \big( \arcsin(1/\sqrt{b}) / 2 \big) - 1 \big]^{-1} \Big) \\
 & - \mathrm{e}^{(b-1)d} \Big[ 2 \arctanh(\sqrt{b}) + \mathrm{i} \pi \big[ 1 + 4 T(\sqrt{2bd},\mathrm{i}/\sqrt{b}) \big] \Big] \Bigg\}
\end{split}
\end{equation}
where $\mathrm{i}^2=-1$, where $T(\chi,c)$ is Owen's T function \cite{Owen1956} defined by
\begin{equation*}
 T(\chi,c) = \frac{1}{2 \pi} \int_0^c \frac{\exp \left[ - \dfrac{1}{2} \chi^2 (1+t^2) \right]}{1+t^2} \id t
\end{equation*}
and where $_2F_2(\{a_1,a_2\},\{b_1,b_2\},\zeta)$ is the generalized hypergeometric function \cite{NIST2010book}.
The functions $T(\sqrt{2bd},\mathrm{i}/\sqrt{b})$ and $_2F_2(\{1,1\},\{3/2,2\},-d)$ can be evaluated in Mathematica using \verb!OwenT[Sqrt[2*b*d],I/Sqrt[b]]! and \verb!HypergeometricPFQ[{1,1},{3/2,2},-d]!, respectively. 

Finally a solvability condition can also be used to show that $L_w=0$, therefore one can substitute $\xi=2/L_0$ in \cref{eq:complicated_integral_velocity}.

\subsection{Self-propulsion driven by capillarity and modulated by activity \label{app:CS_perturbation_solution}}

We consider here the case of pure capillary sliding ($\omega=0$, $\varphi \neq 0$, $\mathcal{A} \neq 0$, $\mathcal{W}=0$), so the governing ODE reduces to
\begin{subequations}
 \label{eq:CS-A_g}
\begin{equation}
  \xi^3 \left( \frac{g^2}{3} + \ell_u g \right) g''' - \mathcal{A} \xi \left( \frac{g}{3} + \ell_u \right) g' = \mathcal{V}
\end{equation}
with boundary conditions
\begin{equation}
\begin{gathered}
 g(\pm 1) = 0, \quad \xi g'(\pm 1) = \mp \phi + \frac{\varphi}{2},
\end{gathered}
\end{equation}
\end{subequations}

If the contact angle are the same ($\varphi=0$), the drop does not move, therefore write the solution to \cref{eq:CS-A_g} as a perturbation power series of both $\varphi$ and $\mathcal{A}$:
\begin{subequations}
\begin{align}
	g & = g_0 + \varphi g_\varphi + \mathcal{A} g_\alpha + \mathcal{A} \varphi g_{\varphi,\alpha} + O(\mathcal{A}^2, \varphi^2) \\
	L & = L_0 + \varphi L_\varphi + \mathcal{A} L_\alpha + \mathcal{A} \varphi L_{\varphi,\alpha} + O(\mathcal{A}^2, \varphi^2) \\
	\mathcal{V} & = \mathcal{V}_0 + \varphi \mathcal{V}_\varphi + \mathcal{A} \mathcal{V}_\alpha + \mathcal{A} \varphi \mathcal{V}_{\varphi,\alpha} + O(\mathcal{A}^2, \varphi^2).
\end{align}
\end{subequations}
The subscript $\varphi$ denotes the correction at order $\varphi$ due to asymmetric contact angles for a passive drop ($\mathcal{A}=0$). It is solution of
\begin{subequations}
\begin{gather}
	g_\varphi''' = \frac{\mathcal{V}_\varphi}{ \xi^3 \left( \dfrac{g_0}{3} + \ell_u \right) g_0} \\
	\text{with\ } \quad g_\varphi(\pm 1) = 0, \quad g_\varphi'(\pm 1) = 1/(2 \xi)
\end{gather}
\end{subequations}
and we find after integration:
\begin{subequations}
\begin{align}
	g_\varphi & = \frac{1}{2 \xi} y + \frac{\beta \phi}{2 \xi (\beta^2-\phi^2)\ln\left[(\beta+\phi)/(\beta-\phi)\right]} \Big\{ \\
	& \left[ (1+y)^2 \ln(1+y)-(1-y)^2\ln(1-y)-2^2 \ln(2) y \right] \nonumber \\
	& + \frac{1}{\beta \phi} \left[ - (\beta+\phi y)^2 \ln(\beta+\phi y) + (\beta-\phi y)^2 \ln(\beta-\phi y) \right] \nonumber \\
	& + \frac{2 y}{\beta} \left[ (\beta+\phi) \ln(\beta+\phi) + (\beta-\phi) \ln(\beta-\phi) \right] \Big\} \nonumber \\
	L_\varphi & = 0 \\
	\mathcal{V}_\varphi & = \frac{\beta \phi}{6 \ln \left( \dfrac{\beta+\phi}{\beta-\phi} \right)}
\end{align}
\end{subequations}
where we have introduced $\beta = \phi \sqrt{1+6 \ell_u \xi/\phi}$.

The subscript $\alpha$ refers to the correction at order $\mathcal{A}$ for a symmetric active drop ($\varphi = 0$, $\mathcal{A} \neq 0$). It is the solution of
\begin{subequations}
\begin{gather}
	g_\alpha''' = \frac{ \mathcal{V}_\alpha + \xi \left( \dfrac{g_0}{3} + \ell_u \right) g_0' }{\xi^3 \left( \dfrac{g_0}{3} + \ell_u \right) g_0 } \\
	\text{with\ } \quad g_\alpha(\pm 1) = 0, \quad g_\alpha'(\pm 1) = 0
\end{gather}
\end{subequations}
and we find
\begin{subequations}
\begin{align}
	g_\alpha & = \frac{1}{2 \xi^2} \big\{ (1+y)^2 \ln(1+y) + (1-y)^2\ln(1-y) \nonumber \\
	& \qquad \qquad \qquad - (2 \ln 2 + 1) y^2 - 2 \ln 2 + 1 \big\} \\
	L_\alpha & = \frac{\Omega}{2 \phi^2} \\
	\mathcal{V}_\alpha & = 0
\end{align}
\end{subequations}

Finally $g_{\varphi,\alpha}$ is the correction at order $\mathcal{A} \varphi$ for an asymmetric active drop which satisfies 
\begin{subequations}
\label{eq:gov_eq_g_alpha_phi}
\begin{gather}
\begin{gathered}
  g_{\varphi,\alpha}''' = \frac{\mathcal{V}_{\varphi,\alpha} - \xi^3 \left( \dfrac{2 g_0}{3} + \ell_u \right) (g_\alpha g_\varphi''' + g_\varphi g_\alpha''')}{\xi^3 \left( \dfrac{g_0}{3} + \ell_u \right) g_0 } \\
  + \frac{\dfrac{1}{3} \xi g_\varphi g_0' + \xi \left( \dfrac{g_0}{3} + \ell_u \right) g_\varphi'}{ \xi^3 \left( \dfrac{g_0}{3} + \ell_u \right) g_0 }
\end{gathered} \\[1em]
	\text{with\ } \quad g_{\varphi,\alpha}(\pm 1) = 0, \qquad g_{\varphi,\alpha}' (\pm 1) = 0
\end{gather}
\end{subequations}
Here we do not solve \cref{eq:gov_eq_g_alpha_phi}, but determine instead $\mathcal{V}_{\varphi,\alpha}$ and $L_{\varphi,\alpha}$ using a solvability condition and we obtain
\begin{subequations}
\begin{align}
	L_{\varphi,\alpha} & = 0 \\
	\mathcal{V}_{\varphi,\alpha} & = \frac{1}{12 \xi \beta} \frac{\beta+\phi}{\beta-\phi} \frac{1}{\left[ \ln \left( \dfrac{\beta+\phi}{\beta-\phi} \right) \right]^2} \Bigg\{ \\
	& - \left[ -2\beta^2 + (\beta-\phi)^2 \right] \dilog \left( \dfrac{\beta+\phi}{\beta-\phi} \right) \nonumber \\
	& - \left[ 2\beta^2 + (\beta-\phi)^2 \right] \dilog \left( \dfrac{2 \phi}{\beta+\phi} \right) \nonumber \\
	& + \beta^2 \left\{ - \left[ \ln(\beta+\phi) \right]^2 + \left[ \ln(\beta-\phi) \right]^2 \right\} \nonumber \\
	& + \left[ 2 \beta^2 + (\beta-\phi)^2 \right] \ln(2\phi) \ln \left( \dfrac{\beta+\phi}{\beta-\phi} \right) \nonumber \\
	& - (\beta-\phi)^2 \left[ 1 + \ln(\beta+\phi) \right] \ln \left( \dfrac{\beta+\phi}{\beta-\phi} \right) \nonumber \\
	& + \frac{\pi^2}{6} \left[ 2 \beta^2 + (\beta-\phi)^2 \right] - 2 \beta \phi \left( \dfrac{\beta-\phi}{\beta+\phi} \right) \Bigg\} \nonumber.
\end{align}
\end{subequations}

Note that since $\mathcal{A}$ affects $L$, it also affects indirectly $\mathcal{V}_\varphi(L)$. We can expand 
\begin{equation}
  \mathcal{V}_\varphi(L)=\mathcal{V}_\varphi(L_0) + \mathcal{A} L_\alpha \frac{\d \mathcal{V}_\varphi}{\d L} \Big\rvert_{L_0}+ O(\mathcal{A}^2,\varphi^2),
\end{equation} 
therefore in the main text we write
\begin{equation}
  \mathcal{V} = \varphi \left( \mathcal{V}_{\varphi,L_0} + \mathcal{A} \mathcal{V}_{\varphi,L_\alpha} \right) + \varphi \mathcal{A} \mathcal{V}_{\varphi,\alpha} + O(\mathcal{A}^2, \tilde{\varphi}^2)
\end{equation}
where $\mathcal{V}_{\varphi,L_0} = \mathcal{V}_\varphi(L_0)$ and $\mathcal{V}_{\varphi,L_\alpha} = L_\alpha \dfrac{\d \mathcal{V}_\varphi}{\d L} \Big\rvert_{L_0}$.

\section{Including nematic stresses \label{app:effect_of_nematic_stresses}}

\begin{figure*}[p]
	\centering
	(a) Self-propulsion driven by active stresses, $\tilde{\mathcal{A}} = (\alpha L)/(2 \pi \omega \gamma \epsilon^2)$, $\tilde{\mathcal{N}} = (K \omega^2 \pi^2)/(4 \gamma L \epsilon^3)$ \\[0.5em]
	\includegraphics[width=0.85\linewidth]{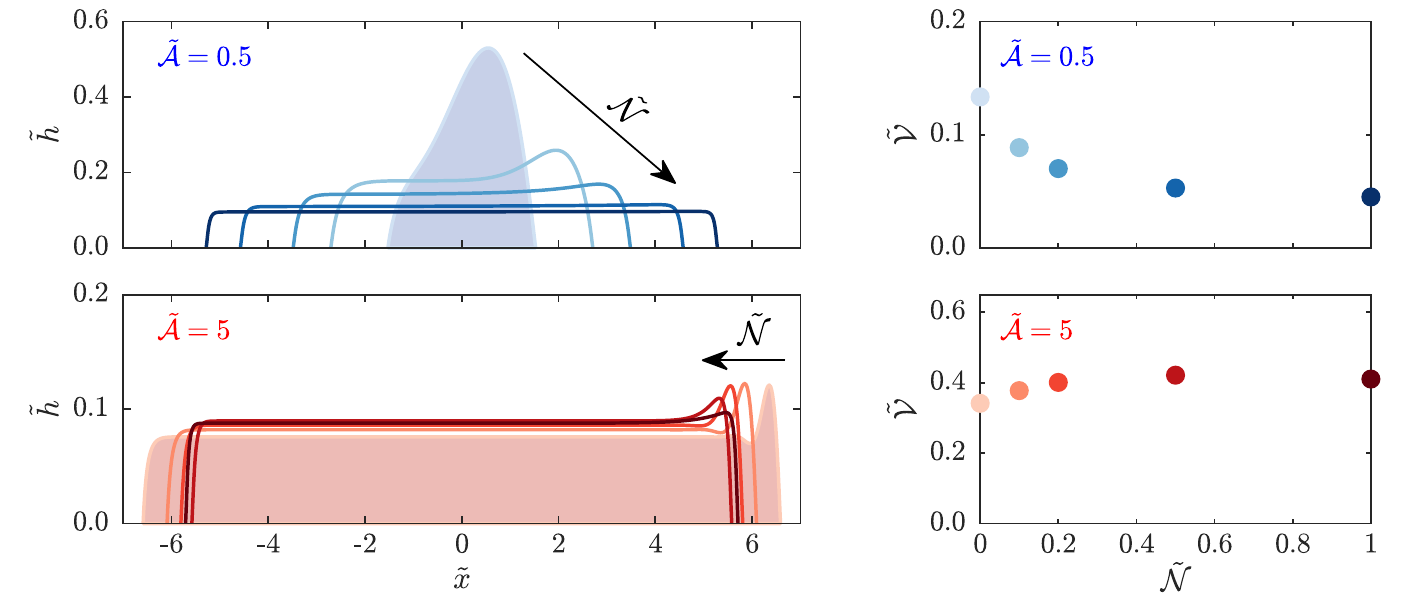} \\[1.5em]
	(b) Self-propulsion driven by self-advection, $\tilde{\mathcal{W}} = (\eta w_0)/(\gamma \epsilon^3)$, $\tilde{\mathcal{N}} = K/(\gamma L \epsilon)$ \\[0.5em]
	\includegraphics[width=0.85\linewidth]{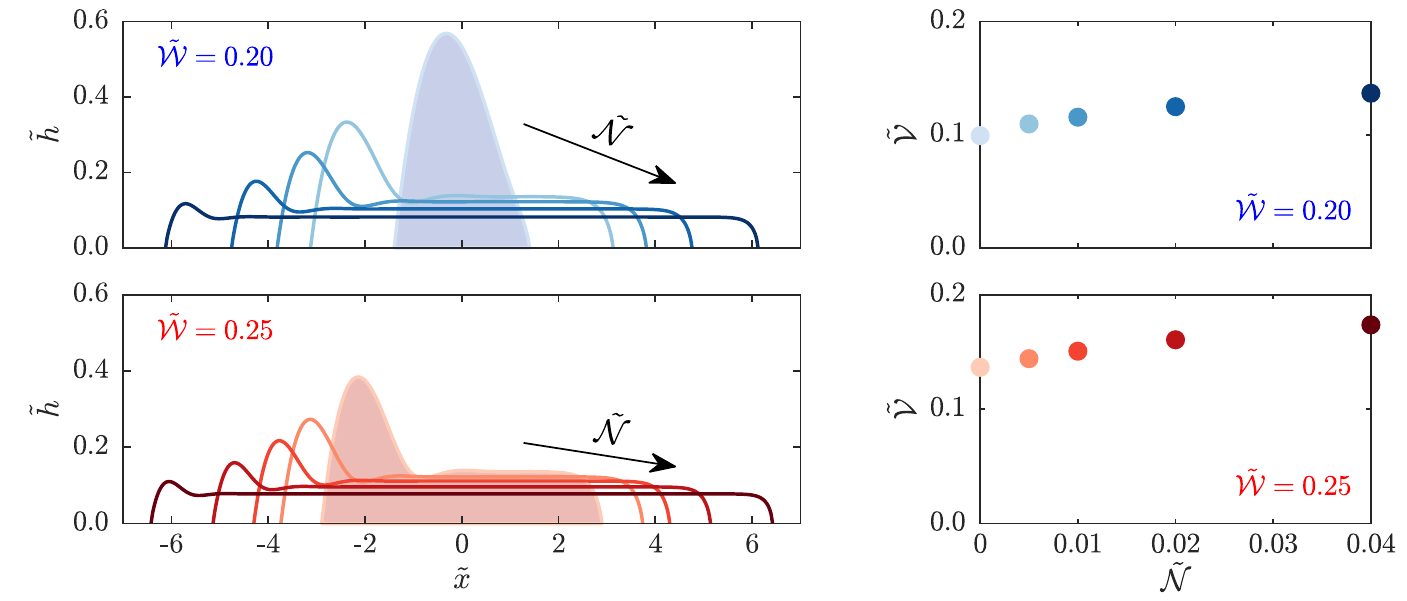} \\[1.5em]
	(c) Self-propulsion driven by capillarity, $\tilde{\mathcal{A}} = (\alpha L)/(\gamma \epsilon)$, $\tilde{\mathcal{N}} = K/(\gamma L \epsilon)$ \\[0.5em]
	\includegraphics[width=0.85\linewidth]{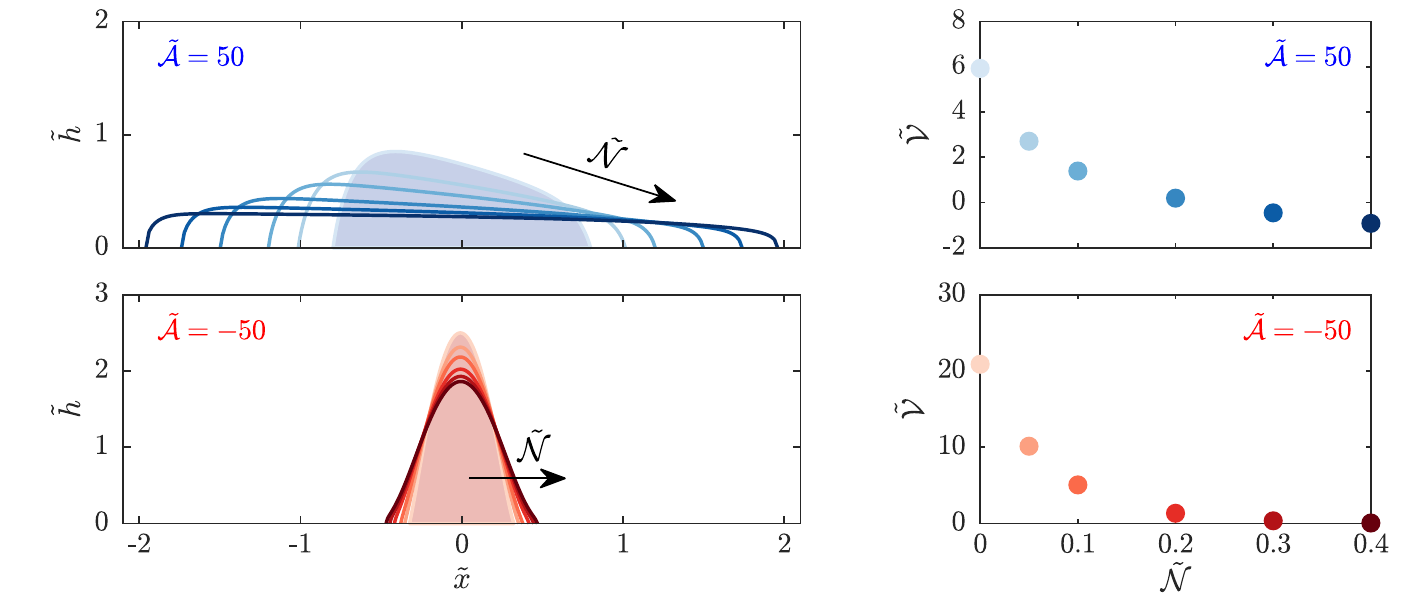} \\
	\caption{
	Effect of nematic stresses on the drop shape and velocity (shaded drops are those obtained for $\tilde{\mathcal{N}}=0$ in the main text), for each mode of motion.
	Colored symbols/lines mark corresponding state points between left and right panels.
	\label{fig:effect_of_nematic_stresses} 
	}
\end{figure*}

We now retain nematic stresses $\sigma_{ij}^n = - K \partial_i n_k \partial_j n_k$ in the momentum balance at leading order. The additional relevant dimensionless parameter is $\mathcal{N}=K/(\eta U L)$. Note that in the following, $m=m(\tilde{h})$ is the regularizing function defined by \cref{eq:m_regularization}, and $m'=\d m/\d \tilde{h}$.

The expressions of the shear stress [\cref{eq:lub_shear_stress}] and of the velocity [\cref{eq;lub_parallel_velocity}] now contain nematic contributions. Those are given by 
% \begin{widetext}
\begin{align}
  \tilde{\sigma}_{xz}^n & =
  \begin{dcases}
  	\frac{\mathcal{N} \pi^2\omega^2}{2 \tilde{h}^3} (m^2 - \tilde{h} m m') \tilde{h}' \left(\tilde{z} - \frac{\tilde{h}}{2} \right) & \text{if $\omega \neq 0$,}\\
  	\frac{2 \mathcal{N} \epsilon^2}{\tilde{h}^3} \left[ (m^2 - \tilde{h} m m') \tilde{h}'^3 - m^2 \tilde{h} \tilde{h}' \tilde{h}'' \right] \left(\tilde{z} - \frac{\tilde{h}}{2} \right) & \text{if $\omega = 0$,}
  \end{dcases} \\
  \tilde{u}_x^n & =
  \begin{dcases}
  	\frac{\mathcal{N} \pi^2\omega^2}{4 \tilde{h}^3} (m^2 - \tilde{h} m m') \tilde{h}' \left(\frac{z^2}{2} - \tilde{h} \tilde{z} -\tilde{\ell}_u \tilde{h} \right) & \text{if $\omega \neq 0$,}\\
  	\frac{\mathcal{N} \epsilon^2}{\tilde{h}^3} \left[ (m^2 - \tilde{h} m m') \tilde{h}'^3 - m^2 \tilde{h} \tilde{h}' \tilde{h}'' \right] \left(\frac{z^2}{2} - \tilde{h} \tilde{z} -\tilde{\ell}_u \tilde{h} \right) & \text{if $\omega = 0$.}
  \end{dcases}
\end{align}
Averaging the velocity over the drop height gives:
\begin{align}
  & \frac{1}{\tilde{h}} \int_0^{\tilde{h}} \tilde{u}_x^n \; \d \tilde{z} = \nonumber \\
  & \begin{dcases}
  	- \frac{\mathcal{N} \omega^2 \pi^2}{4 \tilde{h}^2} \left( \frac{\tilde{h}}{3} + \tilde{\ell}_u \right) (m^2 - \tilde{h} m m') \tilde{h}' & \text{if $\omega \neq 0$,}\\
  	- \frac{\mathcal{N} \epsilon^2}{\tilde{h}^2} \left( \frac{\tilde{h}}{3} + \tilde{\ell}_u \right) \left[ (m^2 - \tilde{h} m m') \tilde{h}'^3 - m^2 \tilde{h} \tilde{h}' \tilde{h}'' \right] & \text{if $\omega = 0$.}
  \end{dcases}
\end{align}
Note that we must have $\mathcal{N} \sim 1$ for $\omega \neq 0$ and $\mathcal{N} \sim \epsilon^{-2}$ for $\omega=0$ such that nematic stresses play a role at leading order.

The thin drop equation (\ref{eq:scaled_ODE}) becomes
\begin{equation}
\begin{split}
  & \left( \frac{\tilde{h}}{3} + \tilde{\ell}_u \right) \tilde{h} \tilde{h}''' + \tilde{\mathcal{A}} \tilde{f}^{\alpha}(\tilde{h}) + \tilde{\mathcal{W}} \tilde{f}^w(\tilde{h}) + \tilde{\mathcal{N}} \tilde{f}^{n}(\tilde{h}) = \tilde{\mathcal{V}} \\
  \tilde{f}^{n}(\tilde{h}) & = 
 \begin{dcases}
 - \frac{1}{\tilde{h}^2} \left( \frac{\tilde{h}}{3} + \tilde{\ell}_u \right) (m^2 - \tilde{h} m m') \tilde{h}'& \text{if $\omega \neq 0$,}\\
 - \frac{1}{\tilde{h}^2} \left( \frac{\tilde{h}}{3} + \tilde{\ell}_u \right) \left[ (m^2 - \tilde{h} m m') \tilde{h}'^3 - m^2 \tilde{h} \tilde{h}' \tilde{h}'' \right]  & \text{if $\omega = 0$,}
\end{dcases}
\end{split}
\end{equation}
with
\begin{equation}
  \tilde{\mathcal{N}} = 
 \begin{dcases}
    \frac{K \omega^2 \pi^2}{4 \gamma L \epsilon^3} & \text{if $\omega \neq 0$,}\\
    \frac{K}{\gamma L \epsilon} & \text{if $\omega = 0$.}
\end{dcases}
\end{equation}

% \end{widetext}

We show in \cref{fig:effect_of_nematic_stresses} how $\tilde{\mathcal{N}}$ affects the solutions presented in the main text. Overall, increasing $\tilde{\mathcal{N}}$ causes the drop to flatten. For self-propulsion driven by active stresses [\cref{fig:effect_of_nematic_stresses}(a)], the tractionless flat solution at high $\tilde{\mathcal{A}}$ is essentially independent of $\tilde{\mathcal{N}}$. For self-propulsion driven by self-advection [\cref{fig:effect_of_nematic_stresses}(b)], $\tilde{\mathcal{N}}$ favors the growth of the protrusion and simply renormalizes the transition to a protruded shape ($\tilde{\mathcal{W}}_c$ decreases upon increasing $\tilde{\mathcal{N}}$). For self-propulsion driven by capillarity [\cref{fig:effect_of_nematic_stresses}(c)], increasing $\tilde{\mathcal{N}}$ dramatically reduces the drop speed.

\section*{Conflicts of interest}
There are no conflicts to declare.

\section*{Acknowledgements}
A.L. thanks Dmitry Zhdanov and Francisco Gonzalez Montoya for deriving (\ref{eq:nasty_integral}). 
Part of this work was funded by a Leverhulme Trust Research Project Grant RPG-2016-147.
T.B.L. acknowledges support of BrisSynBio, a BBSRC/EPSRC Advanced Synthetic Biology Research Centre (grant number BB/L01386X/1).

%%%END OF MAIN TEXT%%%


\balance


%%%REFERENCES%%%
\bibliography{biblio} %You need to replace "rsc" on this line with the name of your .bib file
\bibliographystyle{rsc} %the RSC's .bst file

\end{document}